\documentclass[%
 reprint,
nofootinbib,
 amsmath,amssymb,
 aps,
]{revtex4-2}

\usepackage{graphicx}
\usepackage{dcolumn}
\usepackage{bm}

\usepackage{booktabs}
\usepackage{subcaption}
\usepackage{amsmath}
\usepackage{empheq,etoolbox}
\usepackage{tabularx}
\usepackage{microtype}
\usepackage[utf8]{inputenc}
\usepackage[english]{babel}
\usepackage{siunitx}

\DeclareCaptionJustification{justified}{\leftskip=0pt \rightskip=0pt \parfillskip=0pt plus 1fil}

\patchcmd{\subequations}
  {\theparentequation\alph{equation}}
  {\theparentequation.\alph{equation}}
  {}{}
\DeclareUnicodeCharacter{03B2}{\ensuremath{\beta}}

\begin{document}

\preprint{APS/123-QED}

\title{First Experimental Evidence of a Beam-Beam Long-Range Compensation Using Wires in the Large Hadron Collider}\thanks{Research supported by the HL-LHC project.}

\author{A.~Poyet}
\email{axel.poyet@cern.ch}
\affiliation{CERN, Geneva 1211, Switzerland \\}%
\affiliation{Grenoble-Alpes University, 38400 Saint-Martin-d'Hères, France \\}

\author{A.~Bertarelli, F.~Carra, S.D.~Fartoukh, N.~Fuster-Martínez, N.~Karastathis, Y.~Papaphilippou, M.~Pojer, S.~Redaelli, A.~Rossi, K.~Skoufaris, M.~Solfaroli Camillocci, G.~Sterbini}
\affiliation{CERN, Geneva 1211, Switzerland \\}%

\date{\today}

\begin{abstract}
In high intensity and high energy colliders such as the CERN Large Hadron Collider and its future High Luminosity upgrade, interactions between the two beams around the different Interaction Points impose machine performance limitations. In fact, their effect reduces the beam lifetime and therefore the collider's luminosity reach. Those interactions are called Beam-Beam Long-Range interactions and a possible mitigation of their effect using DC wires was proposed for the first time in the early 2000's. This solution is currently being studied as an option for enhancing the HL-LHC performance. In 2017 and 2018, four demonstrators of wire compensators have been installed in the LHC. A two-year long experimental campaign followed in order to validate the possibility to mitigate the BBLR interactions in the LHC. During this campaign, a proof-of-concept was completed and motivated an additional set of experiments, successfully demonstrating the mitigation of BBLR interactions effects in beam conditions compatible with the operational configuration. This paper reports in detail the preparation of the experimental campaign including the corresponding tracking simulations, the obtained results and draws some perspectives for the future. 
\end{abstract}

\maketitle


\section{Introduction and Motivations}
\label{sec:intro}

In modern particle accelerators, the study of non-linear effects that can possibly have a detrimental impact on the machine performance is of primary importance. In colliders, the interactions between the two beams are part of those limitations. Machines such as the Large Hadron Collider (LHC) \cite{LHC_design} at CERN, or its upgrade, the High-Luminosity LHC (HL-LHC) \cite{HLLHC_design} aim at accelerating and colliding two counter-rotating hadron beams at their Interaction Points (IPs) where massive particle detectors are located. The two beams circulate through a series of superconducting magnets in two different vacuum chambers, separated horizontally by 19.4~cm. Due to the high number of bunches and their short time separation of 25~ns, the two beams collide with a crossing-angle, so that they do not collide outside the detector, which would reduce its efficiency \cite{Sterbini:1229215} and the beam lifetime.

After the recombination dipole, the two vacuum chambers come closer to each other, before turning into a single one before the separation dipole. In between the two separation dipoles, the two beams get closer to each other while approaching the center of the detector. At the IP, the two beams collide in the luminous region \cite{Muratori:691967}, resulting in a Head-On (HO) interaction, responsible for producing the high energy interaction events of interest. The proportionality factor between the event rate and the cross-section of such an event is called instantaneous luminosity (denoted with $\mathcal{L}$) and is typically measured in Hz/cm$^{2}$. Obtaining a closed form for the luminosity is not trivial as different effects have to be taken into account. From \cite{Herr:941318} one can get the expression of the instantaneous luminosity produced by the collisions of two bunches with the same transverse Gaussian distribution ($\sigma_x$, $\sigma_y$) (which is a valid assumption for the LHC and the HL-LHC \cite{Papadopoulou:2631511, Papadopoulou:2275963}), for a given crossing angle $\theta_c$ in the crossing plane and the beams and machine parameters: 
\begin{equation}
\label{eq:lumi_cross}
    \mathcal{L} = \frac{N_1 N_2 f_{\textrm{rev}} N_b}{4 \pi \sigma_x \sigma_y} \frac{1}{\sqrt{1 + (\frac{\sigma_s}{\sigma_x}\tan(\frac{\theta_c}{2}))^2}},
\end{equation}
where $N_i$ ($i=1,2$) is the bunch intensity of the beam $i$, $f_{\textrm{rev}}$ the revolution frequency, $N_b$ the number of colliding bunches at the considered IP, $\sigma_{x,y}$ the RMS transverse beam size at the IP and $\sigma_s$ the RMS bunch length. In Table~\ref{table:lhc_hl_comp} the beam and machine design parameters for the LHC \footnote{The parameters for the LHC are the ones of the original design report. The LHC was operated at an energy of 6.5~TeV during the 2017-2018 Run 2.} and HL-LHC \cite{hl_lhc_baseline} are reported.

\begin{table}[!htbp]
  \caption{Comparison of the LHC (nominal parameters) and HL-LHC baseline parameters.}
  \centering
  \begin{tabularx}{\columnwidth}{Xcc}
    \toprule
    Parameters          &  Nominal LHC          & HL-LHC \\
    \toprule
    Energy [TeV]        &  7                    &  7     \\
    Bunch spacing [ns]  &  25                   &  25    \\
    Number of bunches      &  2808                 &  2760   \\
    Number of collisions (IP1/5)      &  2808                 &  2748   \\
    Bunch population [10$^{11}$]        &  1.15     &  2.2    \\
    Total current [A]   &  0.58                 &  1.11   \\  
    \midrule
    $\beta^{*}$ [cm]    &  55                   &  15 \\
    Full crossing angle $\theta_c$ [µrad] & 285                 &  500 \\
    Beam separation [$\sigma$]  &  9.4          &  10.5      \\
    Normalized transverse emittance [µm]  & 3.75 &  2.5       \\
    \midrule
    Peak luminosity (w/o crab cavities) [10$^{34}$~Hz/cm$^{2}$]  & 1.0  & 8.1 \\
    Leveled luminosity [10$^{34}$~Hz/cm$^{2}$]  & NA  & 5.0 \\
    Leveling time [h] & NA  & 7.2 \\
    \bottomrule
  \end{tabularx}
  \label{table:lhc_hl_comp}
\end{table}

The (HL)-LHC overall performance can be evaluated through the integrated luminosity, typically measured in fb$^{-1}$ (1 barn = 10$^{-24}$~cm$^{2}$), which is the integral over a time interval (generally a year of operation) of the instantaneous luminosity defined previously. Optimizing the integrated luminosity therefore requires to:

\begin{enumerate}
    \item[a.] Improve the bunches overlap at the IP in order to produce the maximum number of collisions,
    \item[b.] Avoid beam losses in order to maintain a good beam lifetime \cite{Crouch:2643365, Pieloni:2294524},
    \item[c.] Increase the integration time by improving the availability of the machine and by reducing the commissioning time.
\end{enumerate}

Concerning (a), one can see from Eq.~(\ref{eq:lumi_cross}) that the instantaneous luminosity can be improved by reducing the crossing angle. However, by doing so, the beam separation is further reduced in the recombination region, in between the two separation dipoles. In the case of proton beams, this would lead to stronger interactions between the two beams as they both create non-linear electromagnetic fields. Those interactions, occurring at a different longitudinal position around the IP, are called Beam-Beam Long-Range (BBLR) interactions. They are one of the main machine performance limitations as they induce beam losses leading to a reduction of the beam lifetime (b) \cite{PhysRevSTAB.2.104001}. As one can see in Table~\ref{table:lhc_hl_comp}, the normalized beam-beam separation in the high-luminosity interaction regions (IR1 and IR5), that is the beam-beam separation given in terms of RMS transverse beam size $\sigma = \sigma_{x,y}$, is 9.4~$\sigma$ for the nominal LHC and 10.5~$\sigma$ for HL-LHC. To recover the luminosity reduction due to a larger crossing angle, the HL-LHC will be equipped with the so-called RF crab cavities \cite{Garlasche:2673651, Calaga:2673544, crab_HL_TDR}. The crab cavities introduce a closed-orbit dependency on the longitudinal position within the bunch, allowing for a maximization of the bunches overlap at the IP. With a 10.5~$\sigma$ beam-beam separation, but a doubled bunch intensity, the effect of the BBLR interactions is expected to be reduced, although not negligible \cite{Skoufaris:2777349}. A proposed solution to cope with the residual BBLR interactions and their possible implications is the use of Direct Current (DC) wires to mitigate the detrimental effect of such phenomena. Despite the specific hardware technical challenges for its alignment \cite{Rossi:2289671, Poyet:2320490}, the use of DC wire compensators is comparable to the other magnets of the machine lattice in terms of operation. Hence, its impact on the machine availability (c) is expected to be reasonable.

The original idea for wire compensators dates back from the early 2000's, by observing the similarity of the BBLR kick with the $1/r$ dependence of a kick given by a DC wire \cite{Papaphilippou:574079, Koutchouk:692058}. The validity of this approximation is improved with large beam-beam separations. After the initial proposal, several experiments have been carried out, installing and testing different types of wires in different machines such as RHIC \cite{PhysRevSTAB.14.091001} at the Brookhaven National Laboratory (BNL), DAFNE \cite{milardi2008dafne} at the INFN in Frascati, Italy and in the SPS at CERN \cite{Zimmermann:1955353, Zimmermann:872262}.

In 2015, a semi-analytical study adopting a resonances compensation criterion to optimize the wires' position and current was carried out \cite{Fartoukh:2052448}, assuming the weak-strong approximation \cite{Pellegrini:2276047, Banfi:1742126}. It was shown first in \cite{Irwin:202623} and then in \cite{Fartoukh:2052448} that the non-linear kicks due to the distributed BBLR interactions could be approximated by two equivalent kicks, one on each side of the IP. This equivalence relies on the fact that with the present (HL)-LHC layout, all the BBLR interactions of a given IR are in phase. The very high $\beta$-functions in the IR ensure an almost constant phase advance on both sides of the IP, while a phase jump of $\pi$ occurs at the IP. All the contributions can therefore be summed in phase and the two equivalent kicks can be computed accordingly. These two kicks can be compensated locally (in terms of phase advance) with DC wires. These wires are assumed to be installed at the same physical transverse distance from the beam, and located at a specific aspect ratio $\beta_x/\beta_y$. It was shown \cite{Fartoukh:2052448} that two wires per IR (one on each side of the IP, installed symmetrically) correctly dimensioned could compensate or minimize all the Resonance Driving Terms (RDT) generated by the BBLR interactions.

This result, together with the fact that the HL-LHC performance might still be limited by the BBLR interactions \cite{Valishev:2013/09/24oga, Karastathis:2715718}, motivated the construction and the installation of wire compensation demonstrators, embedded inside collimators, in the LHC. A two years long experimental campaign followed the installation of the compensators and the main results are reported in this paper. 

The wire demonstrators used in the experiments together with their technical implementation are presented in Section~\ref{sec:exp_setup}. In Section~\ref{sec:exp_protocol}, the observables of the experimental campaign together with the strategies adopted in order to demonstrate the mitigation of the BBLR interactions using wires are discussed. Some of the tracking simulations results \cite{Skoufaris:2777349, Poyet:IPAC19-MOPMP052, poyet_phd} are shown in Section~\ref{sec:simulations}. Finally, the results of the beam measurements campaign are reported in Section \ref{sec:exp_results}.

\section{Setup and hardware layout for the experiment}
\label{sec:exp_setup}

\begin{figure}[!htbp]
\includegraphics[width = 0.9\columnwidth]{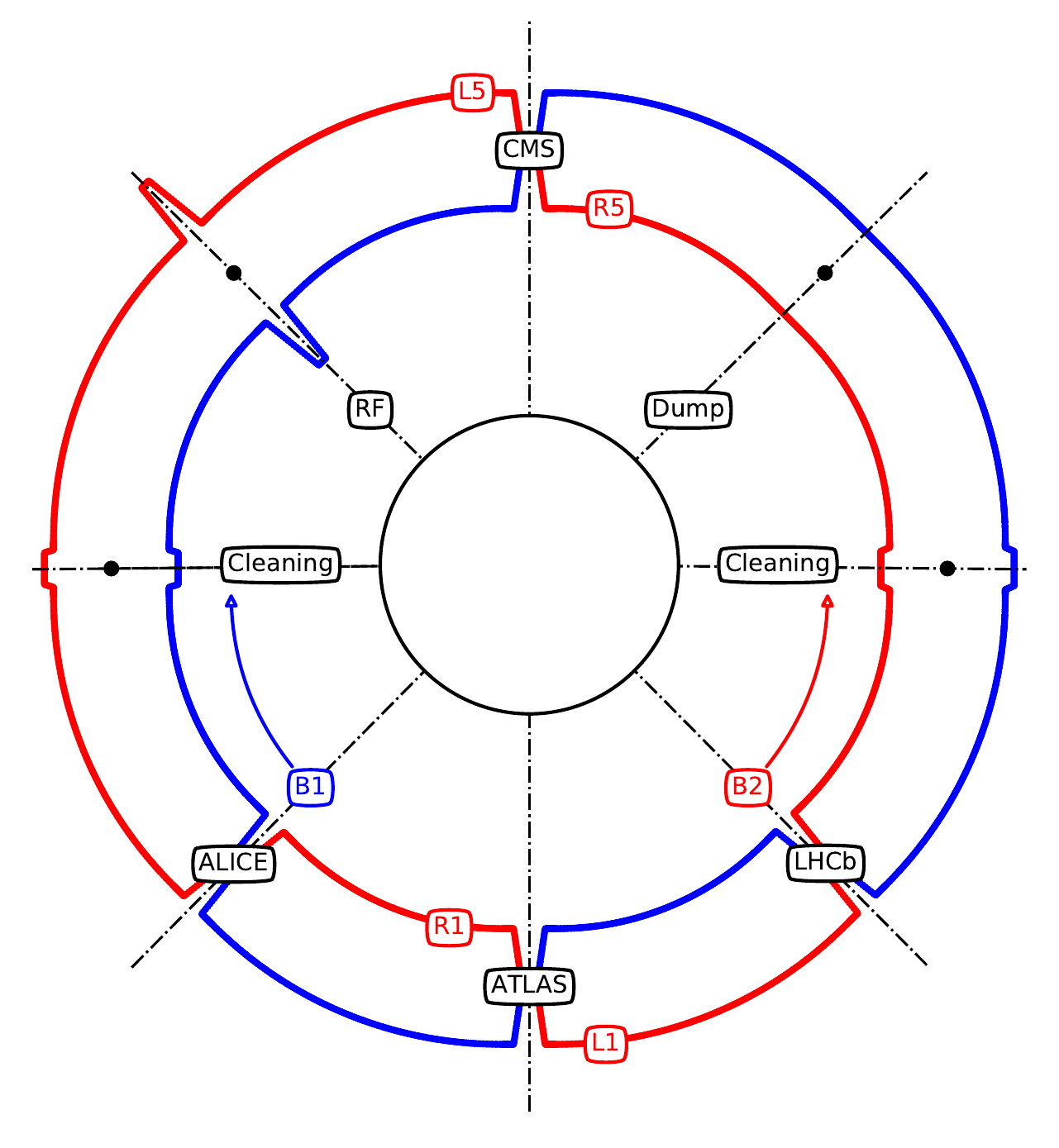} 
\caption{\label{fig:lhc_ring} Out-of-scale schematics of the LHC ring configuration during the 2017-2018 Run. Beam 1 (clockwise) is represented in blue while Beam 2 (anti-clockwise) in red.}
\end{figure}

In the LHC, the two beams collide in four different IPs, around which detectors are installed. Two of those are referred to as high-luminosity IPs, as they host multi-purposed detectors requiring a large number of collisions. Those experiments are ATLAS \cite{ATLAS} in IR1 and CMS \cite{Layter:343814} in IR5 and are diametrically separated in the machine. The crossing angle is rotated by 90$^{\circ}$ between the two IPs: it is vertical in IP1 and horizontal in IP5. The two other experiments are located in the IP 2 and 8, where the ALICE \cite{CERN-LHCC-95-71} and LHCb \cite{CERN-LHCC-95-5} detectors are respectively installed. During the two winter technical stops of 2017 and 2018, demonstrators of BBLR wire compensators have been installed in the LHC \cite{Rossi:2289671}, around the two high-luminosity IPs. The wires are installed for Beam 2 only, as it was the only beam foreseen to operate with a coronograph \cite{Goldblatt:2313940}, which is a device allowing for transverse beam halo measurements. Figure~\ref{fig:lhc_ring} shows the LHC ring together with the different wires, denoted R1, L1, R5 and L5. The naming corresponds to the IP side (Left or Right) and number (IP1 and 5). The longitudinal position of the wires with respect to the nearest IP is given in Table~\ref{table:wire_settings}.

\begin{table}[!htbp]
   \caption{Longitudinal positions of the wires with respect to the corresponding IPs.}
   \centering
   \begin{tabular}{cccc}
       \toprule
       \textbf{Wire} & \textbf{Dist. from IP [m]} & \textbf{Plane} & \textbf{Coll. Name}\\ 
       \midrule
          Wire L1 & -176.17 & V & TCLVW.A5L1.B2\\
          Wire R1 & 145.94 & V & TCTPV.4R1.B2\\
          Wire L5 & -150.03 & H & TCL.4L5.B2\\  
          Wire R5 & 147.94 & H & TCTPH.4R5.B2\\      
       \bottomrule
   \end{tabular}
   \label{table:wire_settings}
\end{table}

\subsection{Magnetic field created by a wire}

Let us consider a DC wire in free space. Since the Cauchy-Riemann conditions are satisfied, the magnetic field created by this wire can be written in the form of a Taylor series \cite{multipolar_cas, Wolski:1333874} as:
\begin{equation}
\label{eq:multipole_wire}
B_y + i B_x = \sum_{n=0}^{\infty} (b_n + i a_n) \frac{(x+iy)^n}{n!},
\end{equation}
where $b_n$ and $a_n$ are the normal and skew components of the $n^{th}$-multipole respectively. We assume the center of expansion (0, 0) to be the reference orbit of the weak beam. The dipolar component of the expansion corresponds to $n=0$. The integrated normal and skew components, denoted $B_n$ and $A_n$ respectively, can be derived from Eq.~(\ref{eq:multipole_wire}) \cite{Fartoukh:2052448}:
\begin{equation}
    \begin{split}
        B_n + i A_n &= \int_{\text{Wire}} b_n + i a_n ds \\
        &= -\frac{\mu_0 (IL)_w}{2 \pi}\frac{n!}{(-x_w - i y_w)^{n+1}},
    \end{split}
    \label{eq:multi_coeff}
\end{equation}
where $(IL)_w$ is the integrated current of the wire expressed in A$\cdot$m, and $(x_w, y_w)$ give the position of the wire with respect to the reference orbit of the weak beam. In MAD-X \cite{madx}, as in the experiment, the weak beam is assumed to be Beam 2, while Beam 1 is the strong beam. A positive integrated current $(IL)_w > 0$ corresponds to a current flowing in the positive s-direction, which is the one of Beam 1.

Using Eq.~(\ref{eq:multi_coeff}), one can get an expression for the normal and skew coefficients using the MAD-X convention. In MAD-X, multipole coefficients are normalized by the beam rigidity $B \rho$ to obtain the normalized and integrated strength of each normal and skew multipole ($KN_n$ and $KS_n$, respectively) as shown below:
\begin{subequations}
  \begin{empheq}[left=\empheqlbrace]{align}
    KN_n &= \frac{B_n}{B\rho}, \\
    KS_n &= \frac{A_n}{B\rho}.
  \end{empheq}
  \label{eq:multi_madx}
\end{subequations}
Finally and using the previous equations, the expression of the $n^{th}$-multipole strengths for a DC wire, is given by:
\begin{subequations}
  \begin{empheq}[left=\empheqlbrace]{align}
    KN_n &= -\frac{n!}{B\rho}\frac{\mu_0 (IL)_w}{2\pi}\Re{\left(\frac{1}{(-x_w-i y_w)^{n+1}} \right)}, \\
    KS_n &= -\frac{n!}{B\rho}\frac{\mu_0 (IL)_w}{2\pi}\Im{\left( \frac{1}{(-x_w-i y_w)^{n+1}} \right)}.
  \end{empheq}
  \label{eq:coef_madx}
\end{subequations}

\subsection{Hardware technical implementation}

\subsubsection{Wires in collimators}
\label{sect:wire_coll}

\begin{figure}[!htbp]
\includegraphics[width = \columnwidth]{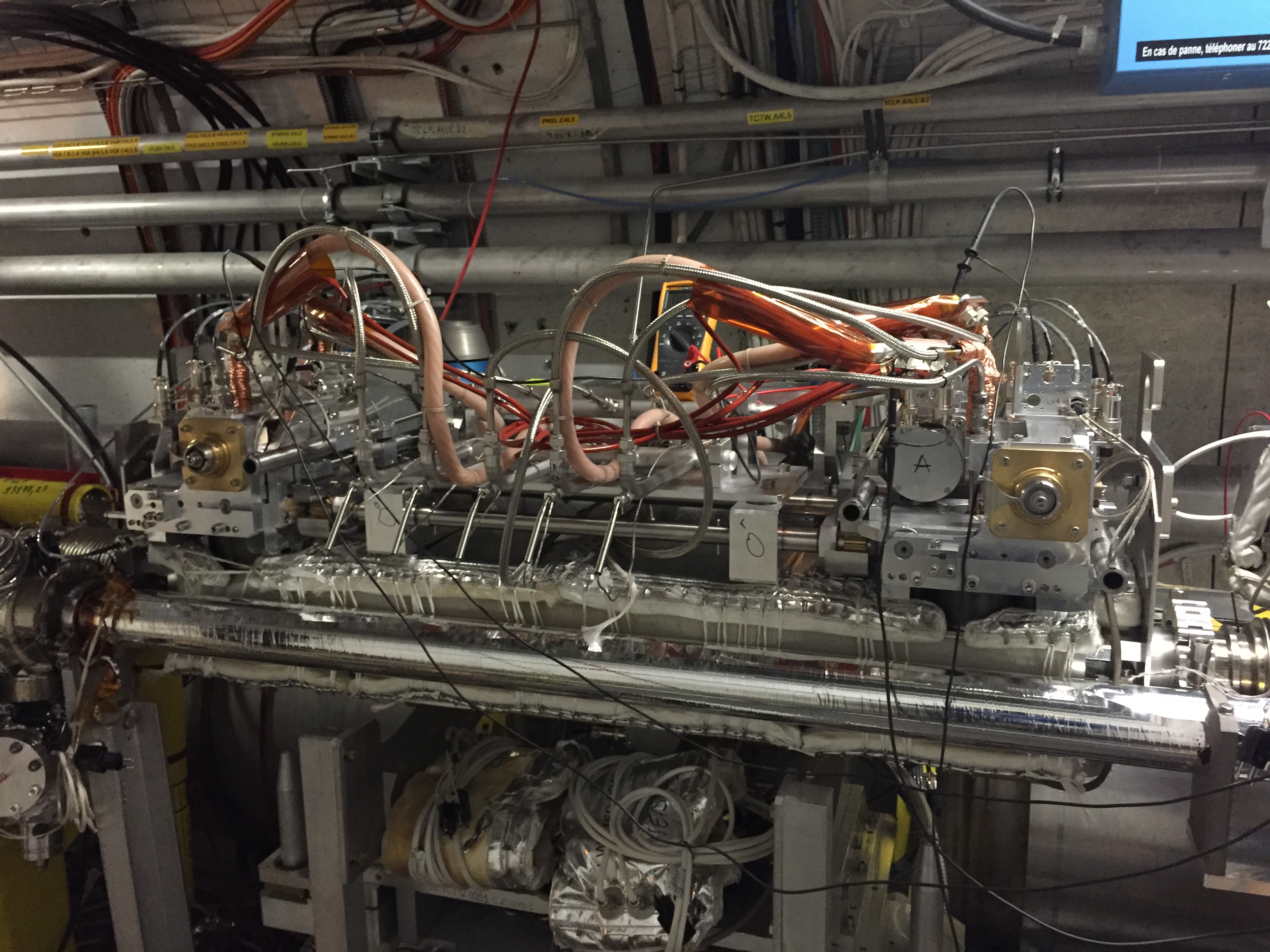} 
\caption{\label{fig:tank} Wire collimator currently installed in the LHC tunnel on the left side of IP5 (L5).}
\end{figure}

Differently from previous wire experiments, the wire demonstrators installed in the LHC are embedded in collimators \cite{Rossi:2289671, Robert-Demolaize:1004869, Assmann:972336}. The LHC collimation system is designed on a multi stages hierarchy. Primary and secondary collimators are located in both IR3 and IR7 of the machine, while tertiary collimators are also located around the different IPs in order to locally protect the Inner Triplet and the experiments \cite{Bruce:2274861}. The wires are housed inside the tungsten jaws of tertiary collimators in the IRs 1 and 5, on Beam 2. Figure \ref{fig:tank} shows one of the four wire collimators currently installed in the LHC tunnel.

Each wire collimator contains two wires, one per jaw. The jaws are water cooled in order to minimize the impact of the beam-induced heating on the overall mechanical structure. The wire moves, together with the housing jaw, inside the vacuum chamber, with an accuracy of about 20~µm. Its transverse position is thus constrained by the collimation hierarchy. The wires should therefore always sit in the shadow of the primary and secondary collimators. Due to the design constraints, the wires' center is located 3~mm behind the jaw of the corresponding collimator. The resulting beam-wire distances are reported in Tables~\ref{table:LI_set} and \ref{table:HI_set} for the different sets of experiments. Finally, the wire collimators have to be aligned with the beam, as a misalignment of the wires would result in a modification of the magnetic field experienced by the beam. The corresponding procedure is reported in Appendix~\ref{app:align}.

\subsubsection{Wire powering configurations}
\label{sect:config}

\begin{figure}[!htbp]
\centering
\begin{subfigure}{.5\columnwidth}
  \centering
  \includegraphics[width=1\linewidth]{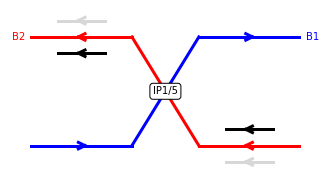}
  \caption{1-jaw powering.}
  \label{fig:dip_config}
\end{subfigure}%
\begin{subfigure}{.5\columnwidth}
  \centering
  \includegraphics[width=1\linewidth]{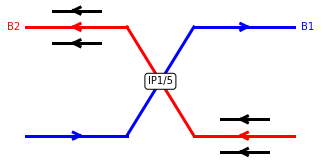}
  \caption{2-jaws powering.}
  \label{fig:quad_config}
\end{subfigure}
\caption{Two different wire configurations were used during the experimental campaign.}
\label{fig:config_wires}
\end{figure}

Each wire can carry up to 350~A, which might not be enough for the compensation, depending on the beam-wire distance. An idea - in order to enhance the wire effect, consists in recabling the two wires of a collimator in series such that they have the same polarity. By doing so, the odd multipolar strengths are doubled while the even ones cancel out. This choice is motivated by the need for a compensation of the octupolar resonances as they represent the first high order effect that is not self-compensated. The two possible configurations are summarized in Fig.~\ref{fig:config_wires}.

\subsubsection{Compensation of the wires linear effects}
\label{sec:FF}

In the LHC, the orbit and tune effects from the BBLR interactions, averaged along the different bunches of the beam, are taken into account for the overall optimization during machine operation \cite{Herr:272882}. The orbit distortion can be compensated using horizontal and vertical orbit correctors. Most of the BBLR-induced tune shift self-compensates as the crossing planes are inverted between the two high luminosity IPs and the optics are symmetric. The wire compensation has to address only the non-linear effects due to the BBLR interactions and the wires linear effects have to be compensated. The LHC closed orbit feedback \cite{Ponce:2294515} is routinely active during the operation of the machine at top energy and is assumed to compensate the wire dipolar effect if the 1-jaw powering configuration is used. This assumption has been verified experimentally \cite{Poyet:2320490}. However, the LHC tune feedback \cite{Ponce:2294515} is not active while the beams are colliding. It is therefore needed to implement a dedicated tune feed-forward to compensate the tune shift induced by the wires as it can reach the $10^{-2}$ level, which is not acceptable for operation.

In order to avoid an important $\beta$-beating wave propagating along the machine, a feed-forward correction is implemented using the nearby quadrupoles for each wire, named Q4 and Q5 \cite{LHC_design}, such that we have: 
\begin{subequations}
  \begin{empheq}[left=\empheqlbrace]{align}
    \Delta Q_{x}^w &= - \Delta Q_{x}^{Q4} - \Delta Q_{x}^{Q5} \\
    \Delta Q_{y}^w &= - \Delta Q_{y}^{Q4} - \Delta Q_{y}^{Q5},
  \end{empheq}
  \label{eq:FF}
\end{subequations}
where $\Delta Q_{x,y}$ represents the horizontal or vertical tune shift induced by the wires (superscript $w$) and by the nearby quadrupoles Q4 or Q5. Assuming each wire to be aligned with the beam, and considering each wire independently, those quantities can be defined as: 
\begin{equation}
    \begin{split}
        \Delta Q_{x,y}^w &= \mp \frac{\mu_0 (IL)_{w}}{8 \pi^2 B \rho} \frac{\beta_{x,y}^w}{d_w^2} \\
        \Delta Q_{x,y}^{Qi} &= \pm \frac{\beta_{x,y}^{Qi} (\Delta KL)_{Qi}}{4 \pi},
    \end{split}
    \label{eq:FF_impl}
\end{equation}
where $d_w$ is the transverse beam-wire distance and $(\Delta KL)_{Qi}$ is the variation of the integrated strength of the quadrupole Q$i$. In Eq.~(\ref{eq:FF_impl}), the 1-jaw powering configuration is assumed. The tune feed-forward can be obtained for the 2-jaws configuration by multiplying the wire induced tune shifts by a factor 2. 

The current LHC uses the so-called Achromatic Telescopic Squeeze (ATS) \cite{Fartoukh:1382077} optics. It consists in using the matching quadrupoles of the considered IR to reach a $\beta^{*}$ ranging from 40~cm to 1~m, depending on the chosen optics while the final $\beta^{*}$ is reached using the matching quadrupoles located in the nearby Interaction Regions. Using the ATS scheme, the luminosity leveling is based on a progressive $\beta^{*}$ reduction along the fill. This $\beta^{*}$-leveling has already been used during the LHC Run 2, will continue to be used during the next LHC Run 3 and is foreseen as baseline for the HL-LHC. During this leveling, the product between $\beta^{*}$ and the $\beta$-functions at any point of the considered IR in between the two Q5 quadrupoles remains almost constant:
\begin{equation}
    \beta_i^{*} \beta(s) \sim \text{Cst}(s), \qquad \forall s \in [\text{Q5.Li}, \text{Q5.Ri}],
\end{equation}
where $i=1,5$ is the IR number. The advantage of this ATS property is that the BBLR compensation problem scales with $\beta^{*}$, consequently requiring a unique feed-forward system during the entire $\beta$-leveling.

The set of Eqs.~\ref{eq:FF} can be applied independently to each wire and their nearby quadrupoles. The goal is then to find a linear relation between the integrated strength variation $(\Delta KL)_{Qi}$ of each quadrupole and the wire parameters $(I_w, d_w)$. We have for each wire:
\begin{subequations}
  \begin{empheq}[left=\empheqlbrace]{align}
    (\Delta KL)_{Q4} &= \alpha_{Q4} \cdot \frac{I_w}{d^2_w} \\
    (\Delta KL)_{Q5} &= \alpha_{Q5} \cdot \frac{I_w}{d^2_w}.
  \end{empheq}
\end{subequations}

Using MAD-X, one can find the $\alpha$ coefficients for the different configurations of the experiment. The $\beta$-beating induced by the wires and their feed-forward system is negligible with respect to the $\beta$-beating induced by the beam-beam effects in the LHC, depending on the chosen optics. More details about the $\beta$-beating study carried out for the wire compensation can be found in Appendix~\ref{app:beta_beat}. These coefficients can also be found analytically from Eq.~(\ref{eq:FF_impl}) in the perturbative approximation, neglecting the residual $\beta$-beating, and the obtained results are consistent with the numerical ones, as the residual $\beta$-beating is negligible.

\section{Experiment's procedure}
\label{sec:exp_protocol} 

\subsection{LHC Typical Cycle and Filling Scheme}

In the LHC, the beams are structured according to a given filling scheme and are generally composed of trains of 144 or 288 bunches \cite{Iadarola:2293539}. In the following, we assume that the two beams collide in IP1 and IP5 only, as the $\beta^*$ and crossing angle are larger in the two other IPs, leading to a negligible effect of the BBLR interactions. With a given beam structure, one can determine the so-called beam-beam collision schedule of a bunch. Depending on its location within a train, a bunch will encounter one or none partner bunches at the IP under the form of a HO collision, and experience a number of BBLR interactions depending on its position. A bunch located in the center of a train will experience more BBLR interactions than a bunch located at the extremity of the same train. 

Once the filling scheme is chosen, the machine is setup for injection, the beams are injected and the energy is ramped from 450~GeV up to 6.5~TeV. The two beams are then squeezed to the desired $\beta^{*}$ and brought into collisions. The BBLR compensation experiments took place at this stage.

\subsection{Observables and objectives of the experiment}
\label{sec:obs_obj}

In an ideal collider, the main mechanism responsible for the beam losses during the collision is the luminosity burn-off: the particles are lost while producing collisions at the interaction point. Considering one IP, the burn-off decay time constant is given by \cite{Antoniou:2293678}: 
\begin{equation}
    \tau_{b0} = \frac{N_0}{\mathcal{L}_0 \sigma_{tot}},
\end{equation}
where $N_0$ and $\mathcal{L}_0$ are the initial bunch intensity and luminosity respectively. $\sigma_{tot}$ is the total proton-proton physical cross-section. It has the dimension of a surface and can be defined, for an ideal collider, as: 
\begin{equation}
    \sigma_{tot} = \sigma_{inel} + \sigma_{el},
\end{equation}
where $\sigma_{el}$ and $\sigma_{inel}$ are the elastic and inelastic cross-section respectively. At 6.5~TeV, $\sigma_{el} \sim$~30~mb and $\sigma_{inel} \sim$~80~mb \cite{Adamczyk:2062926, totem_sigma}. In the LHC, due to the small beam sizes at the IP, only the inelastic part of the proton-proton collisions is expected to contribute to the luminosity losses, the elastic part being mostly responsible for the emittance blow-up. The intensity decay of the considered proton beam can then be defined, for an ideal collider, as: 
\begin{equation}
    N(t) = \frac{N_0}{1+ t/\tau_{b0}}.
\end{equation}
However, in a real machine, beam losses can be caused by different mechanisms, other than the luminosity production. BBLR interactions, electron cloud effects or beam-gas interactions are some examples of sources for beam losses. 

In the LHC, beam losses can be quantified by monitoring the beam intensity evolution. This can be done using the Beam Current Transformers (BCT) \cite{Gras:1300775} or the Fast BCT (FBCT) \cite{Belohrad:1300776}, which provides a bunch-by-bunch intensity measurement. Another way to monitor the beam losses and their location (which is not accessible using the BCT or the FBCT) is the use of the Beam Losses Monitors (BLM) \cite{Garcia}. Those devices give both a spatial and a time resolution of the beam losses. Depending on their bandwidth, one can distinguish the ionisation chamber BLM \cite{ic_blm} which are slower devices and give the averaged losses along the bunches, and the diamond BLM (dBLM) \cite{Gorzawski:2320644}, which are faster and give bunch-by-bunch losses. By differentiating in time the FBCT signal, one can obtain directly the beam intensity loss rate $dN/dt$ while the use of BLMs requires a non-trivial calibration \cite{Salvachua:1574586}. In order to exclude the luminosity losses from the loss rate, one can normalize it to the luminosity $\mathcal{L}_n$ measured in the detector $n$ to obtain the observable of interest for the experiment \cite{Antoniou:2293678}: the effective cross-section, denoted $\sigma_{\textrm{eff}}$. As for the physical cross-section, it has the dimension of a surface and can be defined as:
\begin{equation}
\sigma_{\textrm{eff}} = \frac{1}{\sum_{n \in IP} \mathcal{L}_n} \frac{dN}{dt},
\end{equation}
which can be measured bunch-by-bunch, using the FBCT and the bunch-by-bunch luminosity data provided by the experiments. This observable can be defined only in collisions, with a non-zero luminosity. The drawback of this observable lies on the fact that it requires a precise estimate of the FBCT signal derivative. This can yield a noisy signal and calls for integration periods of several minutes.

\begin{figure}[!h]
\includegraphics[width = \columnwidth]{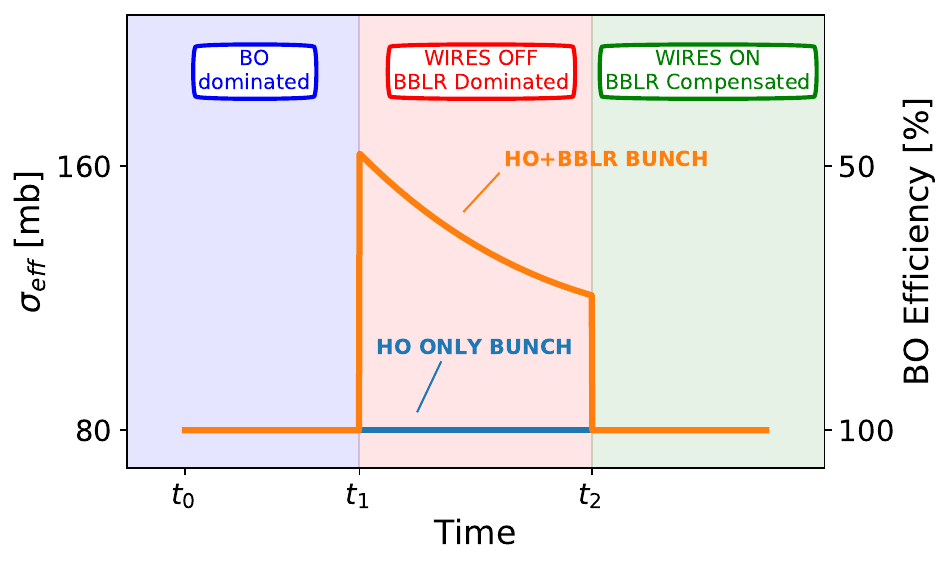} 
\caption{\label{fig:md_rationale} Illustrative example of the effective cross-section and the BO efficiency evolution in time, in an ideal BBLR wire compensation case.}
\end{figure}

In the case of an ideal collider, the effective cross-section is equal to the cross-section of the proton-proton inelastic collisions. However, in presence of other beam loss processes - reducing the beam lifetime, the effective cross-section is overestimated. This allows us to define a burn-off (BO) efficiency, defined as the ratio between the physical and the effective cross-sections, as illustrated in Fig.~\ref{fig:md_rationale}.

If the effective cross-section is measured at $\sim$~80~mb, the beam losses are exclusively due to the luminosity burn-off and the corresponding BO efficiency reaches 100~\%. However, if the effective cross-section increases due to additional losses, the BO efficiency is reduced: in Fig.~\ref{fig:md_rationale}, a cross-section of 160~mb corresponds, in fact, to a burn-off efficiency of 50~\%. 

The objective of the experiment consists in demonstrating that the wire compensators can bring back the BO efficiency to 100~\% in presence of BBLR interactions, by comparing the effective cross-section of two bunches with two different collision schedules: one bunch encountering a partner bunch via a HO collision only (denoted \textit{HO} bunch in Fig.~\ref{fig:md_rationale}) and one bunch encountering additional partner bunches through BBLR interactions (denoted \textit{HO+BBLR} bunch in Fig.~\ref{fig:md_rationale}).

The experiments start in a burn-off dominated regime where the effect of the BBLR interactions is negligible (e.g., with a large crossing angle). At this stage, the effective cross-section of the considered bunches are both equal to $\sim$~80~mb. The effect of BBLR interactions is then enhanced by reducing the crossing-angle or by increasing the transverse beam size by a controlled transverse emittance blow-up \cite{Hofle:1459854}. Only an effect on the effective cross-section on the bunch experiencing the BBLR interactions should be observed under the form of a loss spike followed by a transient. Two diffusion mechanisms are at the origin of this phenomena. The first one consists in interactions between the opposite beam and the high amplitude particles (halo) of the considered one. The BBLR interactions act like a slow beam scraper, inducing losses at first. After some time, as the halo is fully depopulated, the losses decrease and the effective cross-section would return to the 80~mb level. However, and as it will be seen later on with the experimental results, the equilibrium reached after the transient is not 80~mb but a higher value. This indicates a second mechanism in which the BBLR interactions induce a diffusion of the low amplitude particles (core). In this case, the particles diffuse from the core to the halo before being lost. As the core concentrates most of the beam intensity, this explains why the transient does not end at the 80~mb level. Those mechanisms are well understood but remain difficult to quantify and predict. 

Once the BBLR interactions signature is identified, the wires are turned on. If the proof-of-concept is valid, the effective cross-section of the bunch experiencing the BBLR interactions is expected to return to the burn-off level, successfully demonstrating the possibility of mitigating BBLR interactions effects using DC wires. 

\subsection{Beam-wire transverse distance and machine protection}

Depending on the chosen machine configuration, machine protection considerations impose the tertiary collimators opening and, consequently, the beam-wire distances \cite{coll_settings}. Two types of experiments were therefore carried out. A first set of experiment, denoted in the following as Low Intensity (LI) experiment, used lower intensity beams and reduced beam-wire distances. A second set, denoted High Intensity (HI) experiment, used the operational collimators configuration, and it was consequently possible to inject high intensity beams.

\subsubsection{Low Intensity Experiment}

The first set of experiments consisted in a proof-of-concept of a possible mitigation of the BBLR interactions effects using DC wires. In order to enhance the effective strength of the wires, it was requested to close the wire collimators down to 5.5~$\sigma_{\textrm{coll}}$. The subscript ``coll'' refers to the collimation sigma, corresponding to a normalized emittance of 3.5~µm. The corresponding transverse beam-wire distances are reported in Table~\ref{table:LI_set}.

\begin{table}[!h]
   \caption{Wire settings during the Low Intensity experiment. The beam-wire distance is given with respect to the Beam 2 reference system.}
   \centering
   \begin{tabular}{ccc}
       \toprule
       \textbf{Wire} & \textbf{Beam-Wire Distance [mm]} & \textbf{Current [A]}\\ 
       \midrule
          L1 & -7.41 & 350\\
          R1 & 7.42 & 320\\
          L5 & -7.15 & 190\\  
          R5 & 8.24 & 340\\      
       \bottomrule
   \end{tabular}
   \label{table:LI_set}
\end{table}

\begin{figure}[!htbp]
\includegraphics[width = \columnwidth]{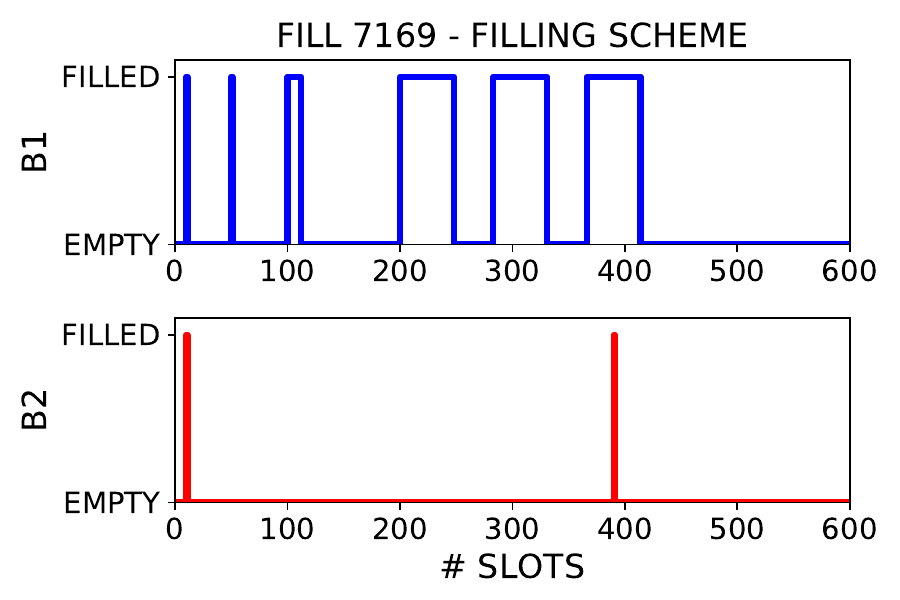} 
\caption{\label{fig:LI_FS} Filling scheme used for the Low Intensity experiment. Only the first 600 bunch slots are displayed.}
\end{figure}

From a machine protection point of view, this required to lower the intensity of the considered beam, namely Beam 2, under the limit of $3 \cdot 10^{11}$ protons. Consequently, it was decided to use the simplest filling scheme providing the wanted collision schedule and only two bunches were used for Beam 2 while Beam 1 could be composed of several trains. Figure~\ref{fig:LI_FS} shows the filling scheme used for the Low Intensity experiment, where the LHC Fill 7169 is taken as an example.

Using this filling scheme, the first bunch of Beam 2 would be the \textit{HO} bunch encountering only one partner bunch via a HO collision. The second one would be the \textit{HO+BBLR} bunch also experiencing BBLR interactions. 

\subsubsection{High Intensity Experiment}

The second set of experiments consisted in observing the possible mitigation of the BBLR interactions effect in a configuration compatible with the nominal LHC operation settings. The collimators opening were therefore set according to the nominal 2018 configuration for physics production and the tertiary collimators were opened at 8.5~$\sigma_{\textrm{coll}}$. The corresponding transverse beam-wire distances are reported in Table~\ref{table:HI_set}.

\begin{table}[!htbp]
   \caption{Wire settings during the High Intensity experiment. The beam-wire distance is given with respect to the Beam 2 reference system. The mention ``x 2'' indicates that the 2-jaw powering configuration is used.}
   \centering
   \begin{tabular}{ccc}
       \toprule
       \textbf{Wire} & \textbf{Beam-Wire Distance [mm]} & \textbf{Current [A]}\\ 
       \midrule
          R1 & 9.83 & 350 x 2\\
          R5 & 11.10 & 350 x 2\\      
       \bottomrule
   \end{tabular}
   \label{table:HI_set}
\end{table}

Only three out of the four wire collimators are operational. The collimator L1 is, in fact, not a tertiary collimator. During the nominal operation of the LHC, its jaws have to sit at a normalized distance larger than 15~$\sigma_{coll}$. It was decided not to use it as the expected effect of wires located at such distance is weak. The L5 wire collimator is a tertiary collimator and is therefore operational. However, in order to maintain the symmetry between the two IPs, it was decided not to use the corresponding wires. 

\begin{figure}
\includegraphics[width = \columnwidth]{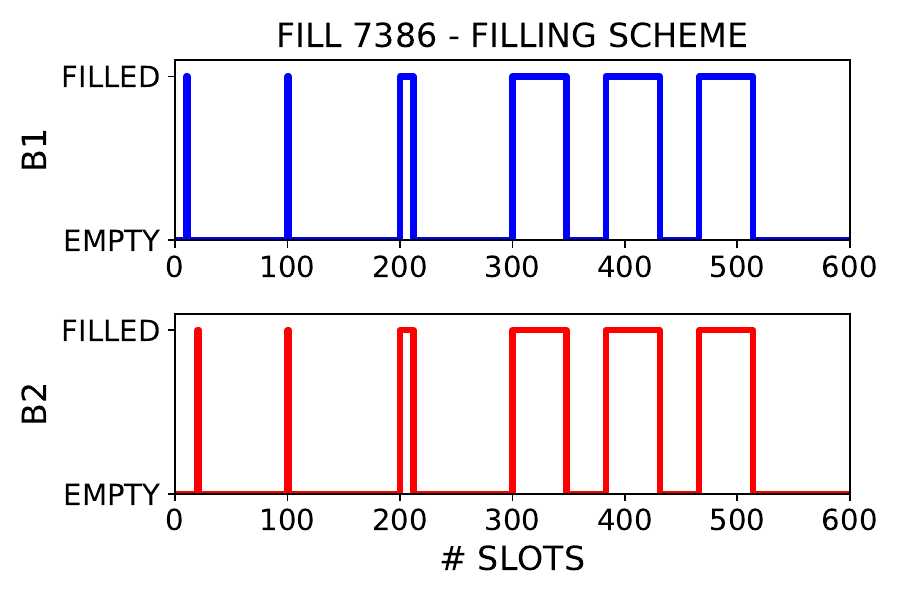} 
\caption{\label{fig:HI_FS} Filling scheme used for the High Intensity experiment. Only the first 600 bunch slots are displayed.}
\end{figure}

With such collimator settings, it was possible, from a machine protection point of view, to use a higher intensity for Beam 2, injecting several trains of bunches as for Beam 1.  Figure~\ref{fig:HI_FS} shows the filling scheme used for the High Intensity experiment, where the LHC Fill 7386 is taken as an example.

\subsection{Wire currents settings}
\label{sec:wire_settings}

\begin{figure}[!h]
\includegraphics[width = \columnwidth]{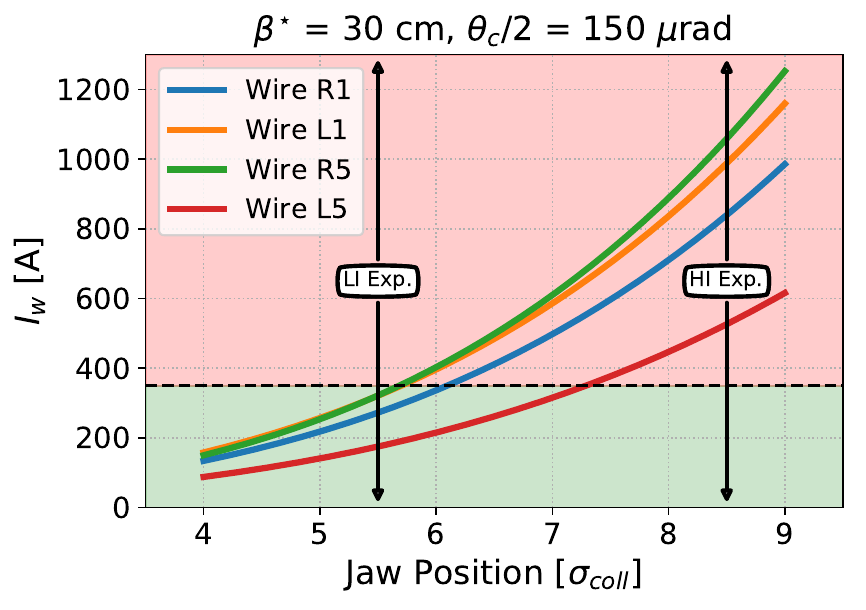}
\caption{\label{fig:opt_currents} Wire currents required to compensate the (4,0)-(0,4) RDTs.}
\end{figure}

Since the transverse beam-wire distance is fixed by the collimators, the wire currents are left as the only adjustable settings. The choice of the wire currents was inspired by the rationale of \cite{Fartoukh:2052448}, and adapted according to the technical constraints of the present hardware layout. Even if an optimal compensation of all the Resonance Driving Terms cannot be reached, numerical studies showed that with the present configuration, compensating the (4,0)-(0,4) RDTs would lead to a partial compensation of all the others. Those RDTs correspond to the octupolar resonances, playing a major role in the BBLR interactions and their mitigation process. Using the settings of the experiment, one can compute the optimal currents needed in order to compensate the (4,0)-(0,4) RDTs as a function of the transverse beam-wire distance. The analytical computation is reported in Appendix~\ref{app:rdt_comp} and the optimal currents, as a function of the wire collimator opening, are reported in Fig.~\ref{fig:opt_currents}.

In the case of the Low Intensity experiment, the needed currents were within the possible reach of 350~A in each wire and the 1-jaw powering could be used. However, it was not the case for the High Intensity experiment as shown in Fig.~\ref{fig:opt_currents}, and the currents were pushed to the limits using the 2-jaws powering configuration (Fig.~\ref{fig:config_wires}), as it would double the strength of the wire octupolar field. The exact values of the wire currents used in the different experiments are reported in the Tables~\ref{table:LI_set} and \ref{table:HI_set}.

\section{Numerical Simulations}
\label{sec:simulations}

The dimensioning of the wire compensators was also supported by numerous tracking simulations, whose results are reported in details in \cite{Skoufaris:2777349, Poyet:IPAC19-MOPMP052, poyet_phd}. In this section, we report part of those results, focusing on the effect of the wire currents and the beam-wire distance on the Dynamic Aperture (DA). The choice of this observable comes from its demonstrated link with the beam lifetime \cite{PhysRevSTAB.15.024001}, making it a good way to anticipate the performance of the machine. The tracking studies are done using both MAD-X and SixTrack \cite{sixtrack}. Simulations are performed tracking particles over a million turns, and - unless specify otherwise, following the parameters given in Table~\ref{table:simu_parameters}.

\begin{table}[!hbt]
   \centering
   \caption{Simulation parameters}
   \begin{tabular}{lcc}
       \toprule
       \textbf{Parameter} & \textbf{Symbol} & \textbf{Reference value}                       \\
       \midrule
           Bunch Intensity                      & $N_b$                   & \SI{1.15E11}{p}                   \\ 
           $\beta$-function at the IP           & $\beta^{\ast}$          & \SI{30}{cm}                       \\
           Half crossing-angle                  & $\theta_c/2$            & \SI{150}{µrad}                \\
           Tunes                                & $Q_x, Q_y$              & 62.31, 60.32                      \\ 
           Chromaticities                       & $\xi_{x,y}$             & 15                                \\ 
           Octupole Current                     & $I_{MO}$                & \SI{0}{A}                         \\
           Number of turns                      &                         & 10$^{6}$                          \\
       \bottomrule
   \end{tabular}
   \label{table:simu_parameters}
   
\end{table}

\begin{figure}[!h]
    \includegraphics[width = \columnwidth]{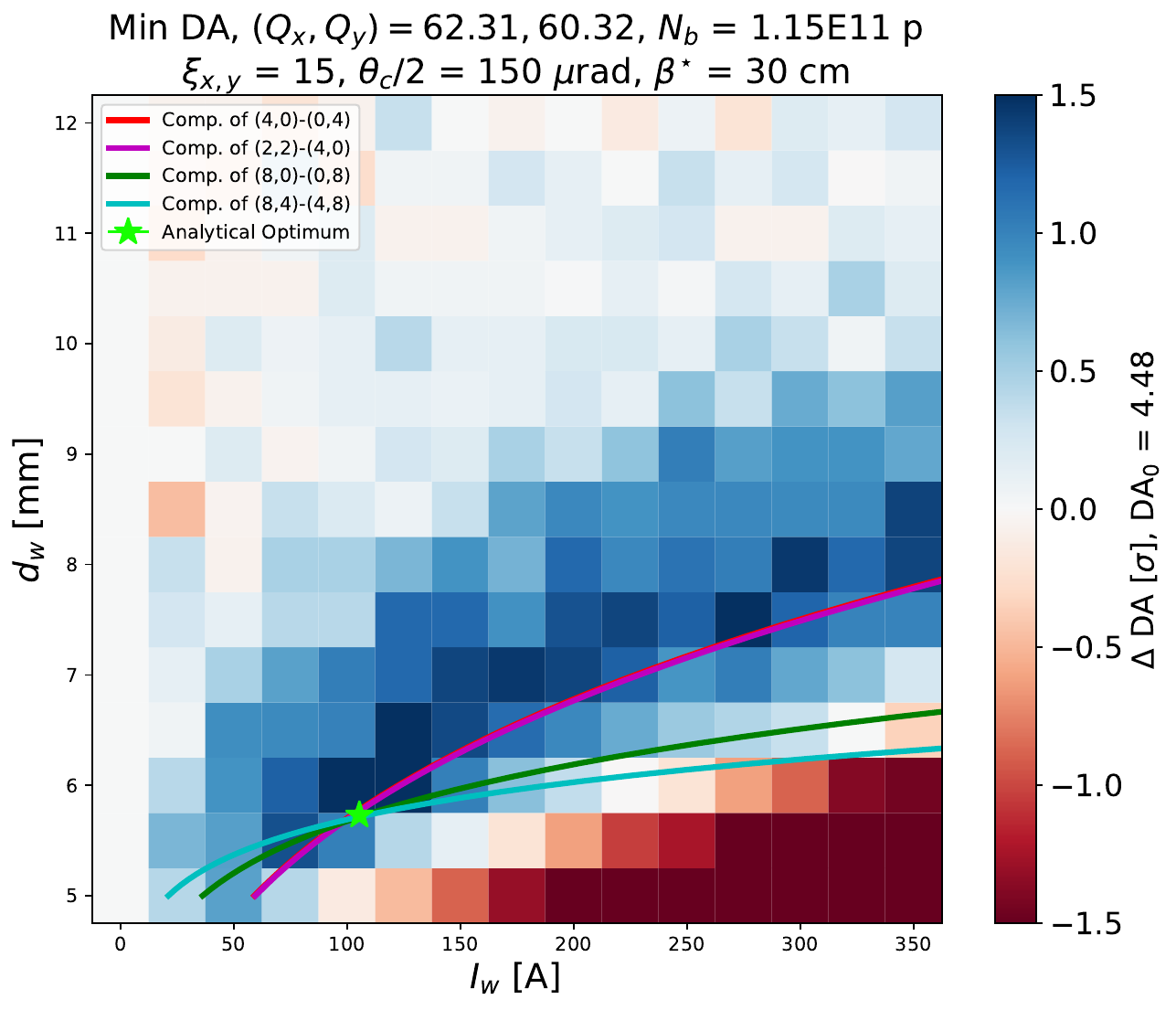} 
    \caption{\label{fig:da_ideal} DA variation (from red for a loss of DA to blue for a gain) as a function of the beam-wire distance and the wires current for the analytical case. The different colored lines show the configurations needed to compensate a given RDT.}
\end{figure}

As discussed in Section~\ref{sec:exp_protocol}, the first dimensioning was done using an RDT-compensation approach, based on the work presented in \cite{Fartoukh:2052448}. In that paper, several assumptions were made: all the wires are located at the same distance from the IP (ensuring the same $\beta$-aspect ratio for all them, the physical beam-wire distance is equal for all the wire, and all the wires are powered with the same current. In Fig.~\ref{fig:da_ideal}, we show the dependency of the DA on the beam-wire distance $d_w$ and the wires current $I_w$, in this ideal configuration. 

The DA is represented as a variation from a configuration without wires. The red color shows a degradation of the DA while the blue color represents an improvement. As expected from \cite{Fartoukh:2052448}, all the RDT are compensated for a given wire configuration (showed with the green star): all the RDT-compensation lines (colored plain lines) cross in one single point. However, this study also shows that - in terms of DA - the analytical optimum is not the only possible configuration that would lead to an improvement of the machine performance. Instead, we can observe a large dark blue area, following the (4,0)-(0,4) and (2,2)-(4,0) RDT lines. This confirms that the octupolar resonances are the ones to be targeted by the wire compensators. Finally, installing an object so close from the beam (about 5-6~mm) would not be reasonable from a machine protection point of view, and this study therefore confirmed that it is possible to put the wires further away, by powering them with higher currents. 

\begin{figure}[!h]
    \includegraphics[width = \columnwidth]{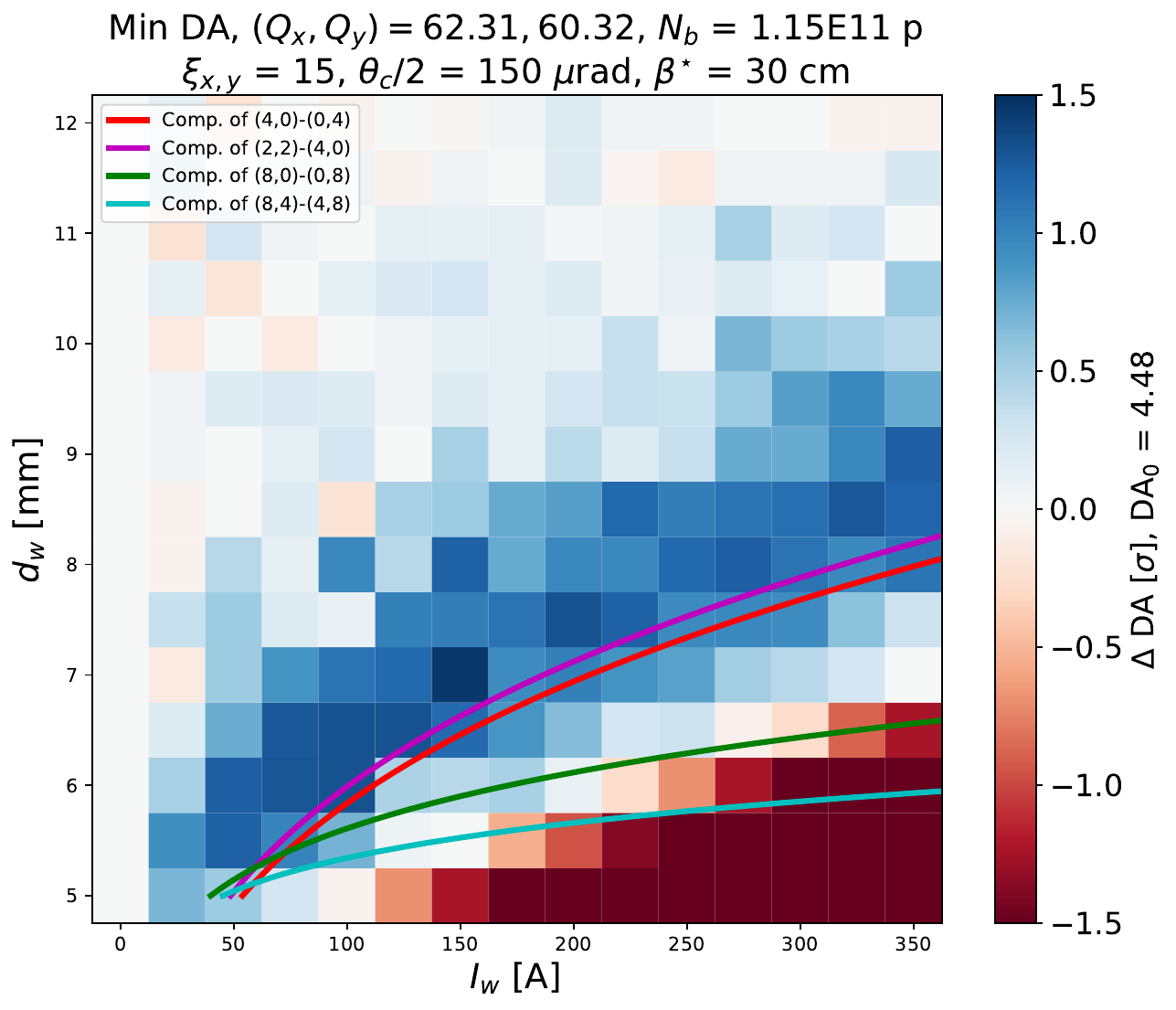} 
    \caption{\label{fig:real_da} DA variation (from red for a loss of DA to blue for a gain) as a function of the beam-wire distance and the wires current for the actual LHC case. The different colored lines show the configurations needed to compensate a given RDT.}
\end{figure}

We then repeated this study for the actual LHC, where the longitudinal positioning of the wires is not ideal anymore (see Table~\ref{table:wire_settings}). Figure~\ref{fig:real_da} shows the results of this study. 

This time, the RDT-compensation lines do not cross anymore, showing that a perfect compensation is not achievable. However, the two other observations remain: it is still possible to put the wires further away from the beam by powering them with higher currents, and the DA improvement is well correlated with the compensation of the octupolar RDTs. 

\begin{figure}[!h]
    \includegraphics[width = \columnwidth]{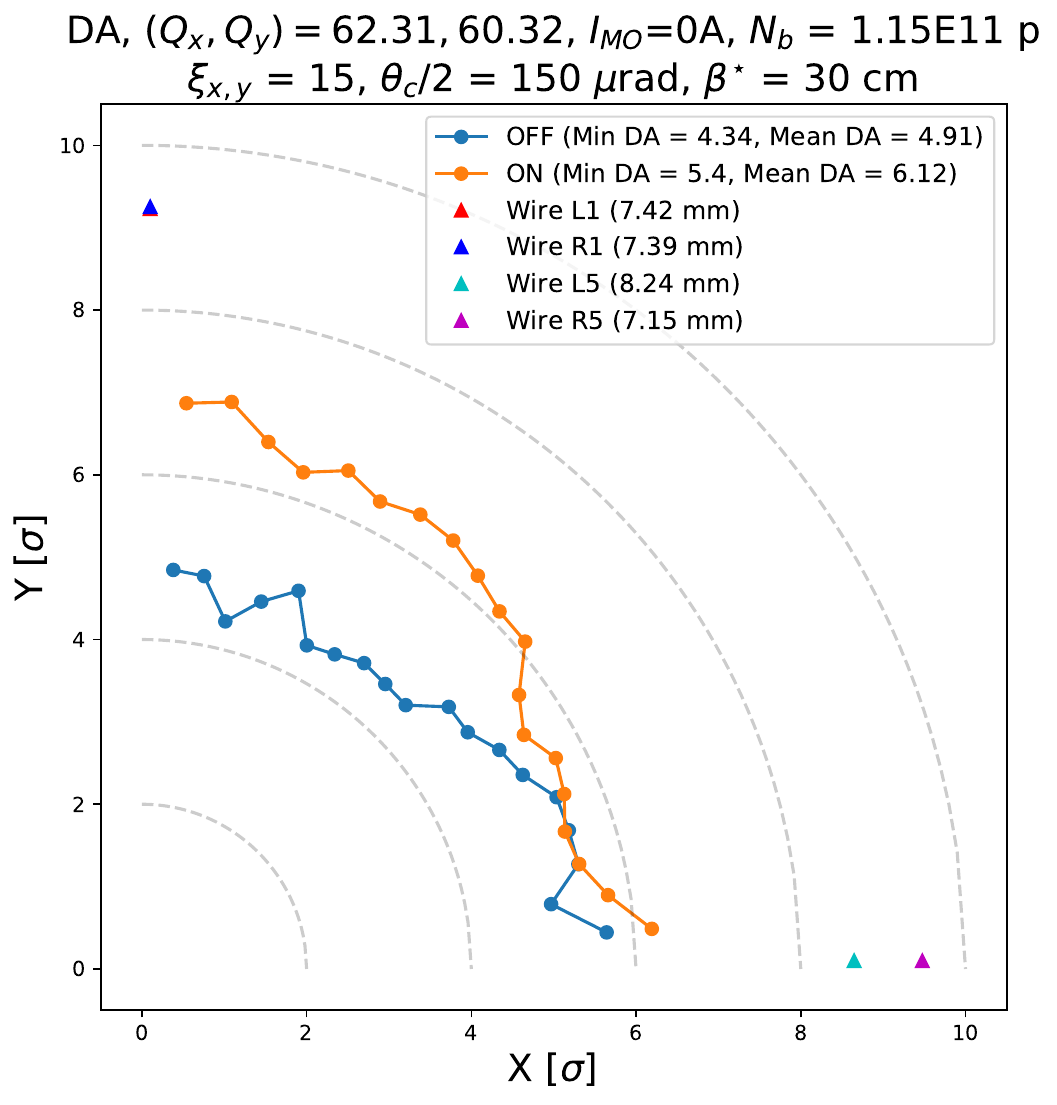} 
    \caption{\label{fig:da_angle} DA in the configuration with the wires ON (orange) and OFF (blue). The wires transverse position is depicted by colored triangles.}
\end{figure}

Using this simulation, and relaxing the assumption that all wires should be placed at the same physical transverse distance from the beam (by necessity, from the collimators) and powered with the same currents (by choice, from the analytical considerations showed in Section~\ref{sec:exp_protocol}), we studied the impact of the wire compensators on the DA in the exact Low Intensity configuration. The results are reported in Figure~\ref{fig:da_angle}.

We observed a mean DA improvement of 1.21~$\sigma$. The improvement seems to be more important in the vertical plane, although this observation is not fully understood yet. 

\begin{figure*}
\centering
\begin{subfigure}{1\columnwidth}
  \centering
  \includegraphics[width=1\linewidth]{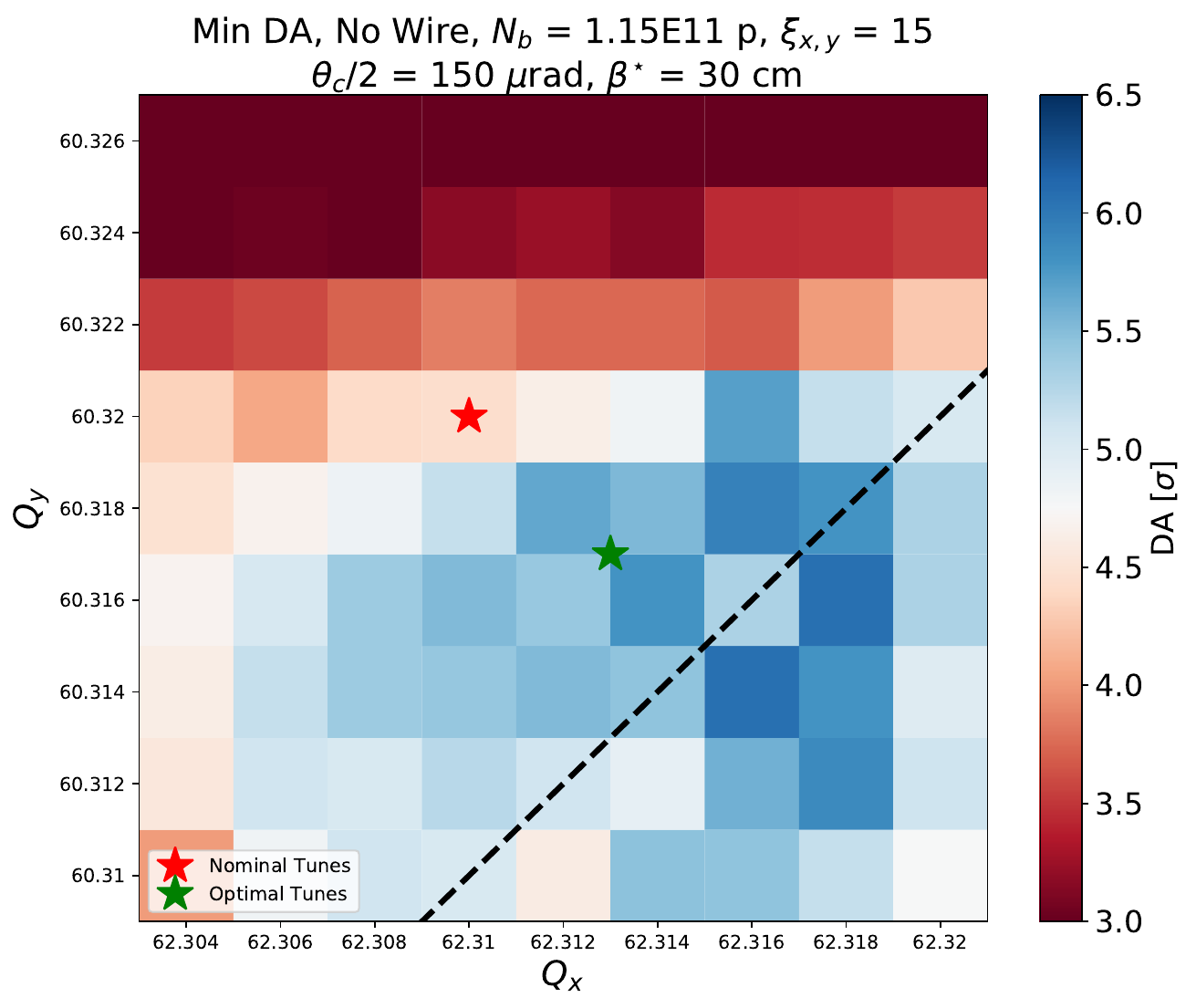}
  \caption{Wires OFF.}
\end{subfigure}%
\begin{subfigure}{1\columnwidth}
  \centering
  \includegraphics[width=1\linewidth]{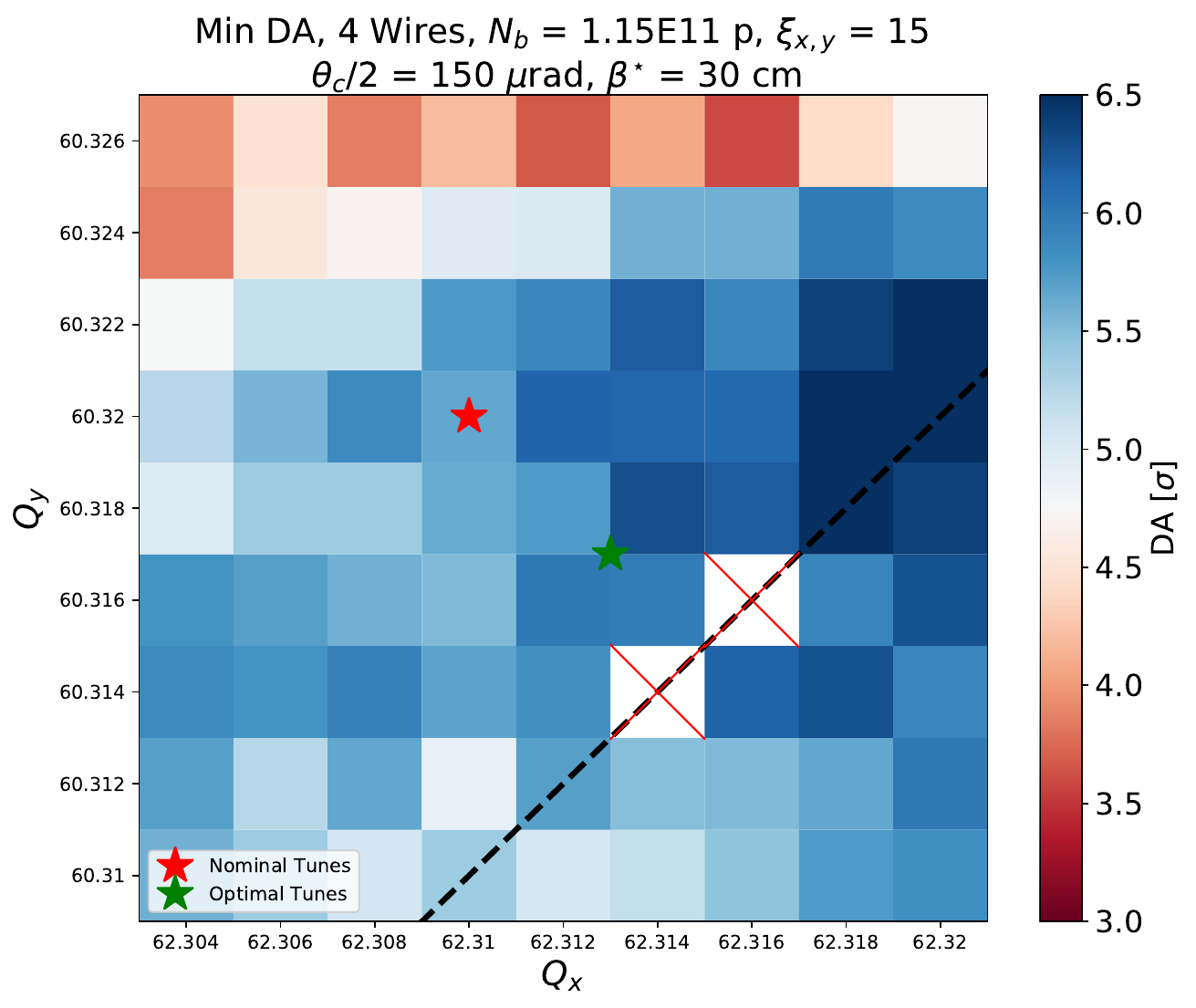}
  \caption{Wires ON.}
\end{subfigure}
\caption{Dependency of the DA on the horizontal and vertical tunes, with (left) and without (right) wires. The red star shows the nominal LHC tunes, while the green star shows the optimal tunes usually used in operation.}
\label{fig:tune_scan_da}
\end{figure*}

Finally, the beneficial impact of the wire compensators on the machine performance can be observed through tune scans, both with and without wires. This study result is reported in Fig.~\ref{fig:tune_scan_da}.

From those studies, we can observe that the wire compensators allow for an opening of the tune space in terms of DA, especially towards the diagonal and the third-integer resonance. This could also help accommodate detrimental effects coming from non-linear phenomena such as electron clouds \cite{Zimmermann:585562, Paraschou:2696125}. 

To conclude, those studies confirmed the potential of the wire compensators in the LHC and led to the experimental campaign, whose results are reported in the next Section.

\section{ Experimental Results and Observations}
\label{sec:exp_results}

Finally we report the results from the Low and High Intensity experiments in this section. More detailed results can be found in \cite{Sterbini:2703495,Poyet:2703503}.

\subsection{Low Intensity Experiment}

Using the settings reported in Table~\ref{table:LI_set}, the Low Intensity experiment was carried out following the previously described procedure. The polarity of the wires was checked, together with their correct alignment with the beam. The starting point in collisions was chosen at a crossing angle of 150~µrad (corresponding to a normalized crossing angle of about 8.6~$\sigma$) and a $\beta^{*}$ of 30~cm.

\begin{figure*}[!htbp]
\includegraphics[width = 2\columnwidth]{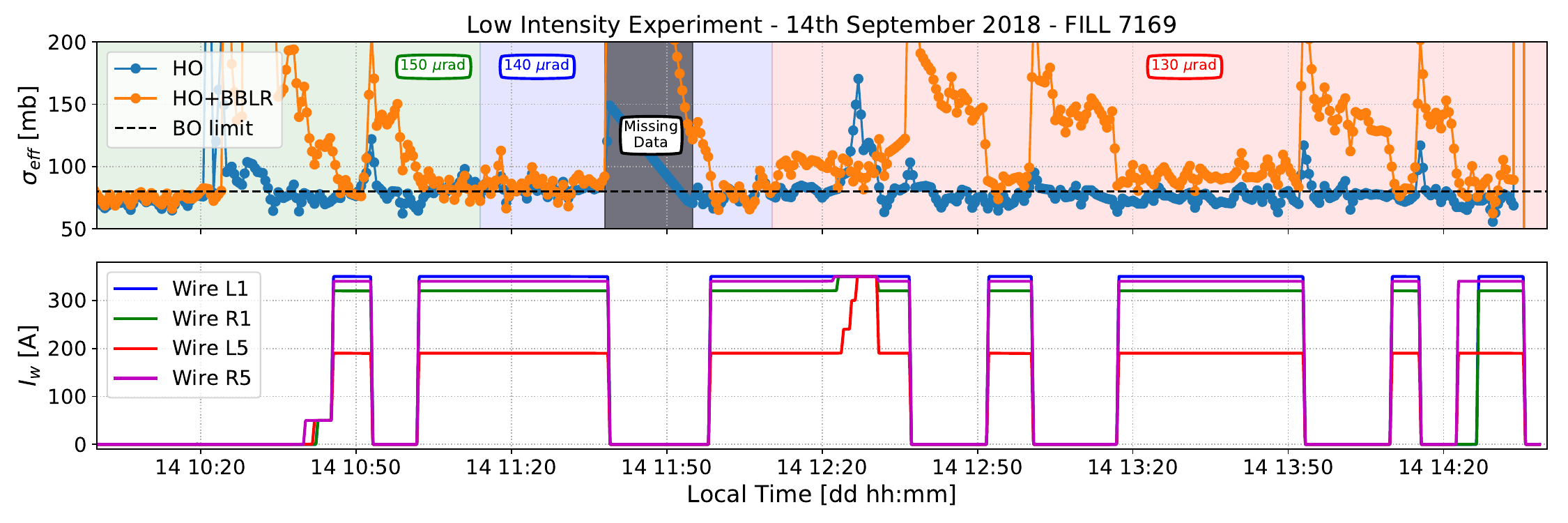} 
\caption{\label{fig:LI_results} Low Intensity experiment results. The top plot shows the evolution of the effective cross-section for both the \textit{HO} bunch (in blue) and the \textit{HO+BBLR} bunch (in orange), together with the burn-off limit (black dashed line). The background coloring corresponds to the different values of crossing angle. The bottom plot shows the evolution of the current in the four different wires.}
\end{figure*}

In order to identify the BBLR interactions signature, the transverse beam size of Beam 2 was increased by a controlled transverse emittance blow-up. A spike of beam losses was observed on the \textit{HO+BBLR} indicating that the BBLR interactions have been enhanced. The wire compensators were thus turned on and off repeatedly in cycles, and the results are reported in Fig.~\ref{fig:LI_results}. 

The top plot on Fig.~\ref{fig:LI_results} shows the evolution of the effective cross-section for both the \textit{HO} (in blue) and \textit{HO+BBLR} (in orange) bunches, together with the burn-ff limit (black dashed line). The background coloring corresponds to the different values of crossing angle. The bottom plot shows the evolution of the current in the four different wires. The spike in the effective cross-section observed around 10:20 corresponds to the controlled transverse emittance blow-up. The \textit{HO} bunch also experienced a spike of a smaller amplitude. By powering on and off the wires, the losses of the \textit{HO+BBLR} bunch reduced and increased respectively, showing, with a crossing angle of 150~µrad, a possible mitigation of the BBLR interactions with the wires. At around 11:10, the crossing angle was reduced down to 140~$\mu$rad (8.1~$\sigma$), with the wires on, enhancing again the BBLR interactions. Nevertheless, no spike of losses was observed for the \textit{HO+BBLR} bunch, showing that even with a reduced crossing angle, the wire compensators could still be effective. The crossing angle was finally decreased further down to 130~µrad (7.5~$\sigma$). A slight increase of losses was observed for the \textit{HO+BBLR} bunch but the effective cross-section remained close to the burn-off limit. Eventually, some on and off cycles with the wires were performed in order to demonstrate the reproducibility of the observations, even in a more aggressive configuration with a lower crossing angle. At around 11:50, data was partially lost due to an issue with the CMS data logging system. Finally, at 12:25, all the wire currents were set to 350~A, and a loss spike was observed for the bunches suffering only from HO collisions. This observation is compatible with an overcompensation mechanism: the wires compensators act like additional BBLR interactions inducing beam losses.

The Low Intensity experiment was therefore very conclusive, showing a clear effect of the wire compensators on the BBLR interactions. It has been observed that the BBLR-induced losses could be reduced or cancelled by powering the wires, without any negative impact on the bunch experiencing no or less BBLR interactions. Those positive results motivated the High Intensity experiment, moving towards beam conditions compatible with the nominal operation configuration.

\subsection{High Intensity Experiment}

The High Intensity experiment was carried out using a similar procedure and the wire settings are recalled in Table~\ref{table:HI_set}. The starting crossing angle for the collision was 160~µrad (9.2~$\sigma$) with a $\beta^{*}$ of 30~cm.

\begin{figure*}[!htbp]
\includegraphics[width = 2\columnwidth]{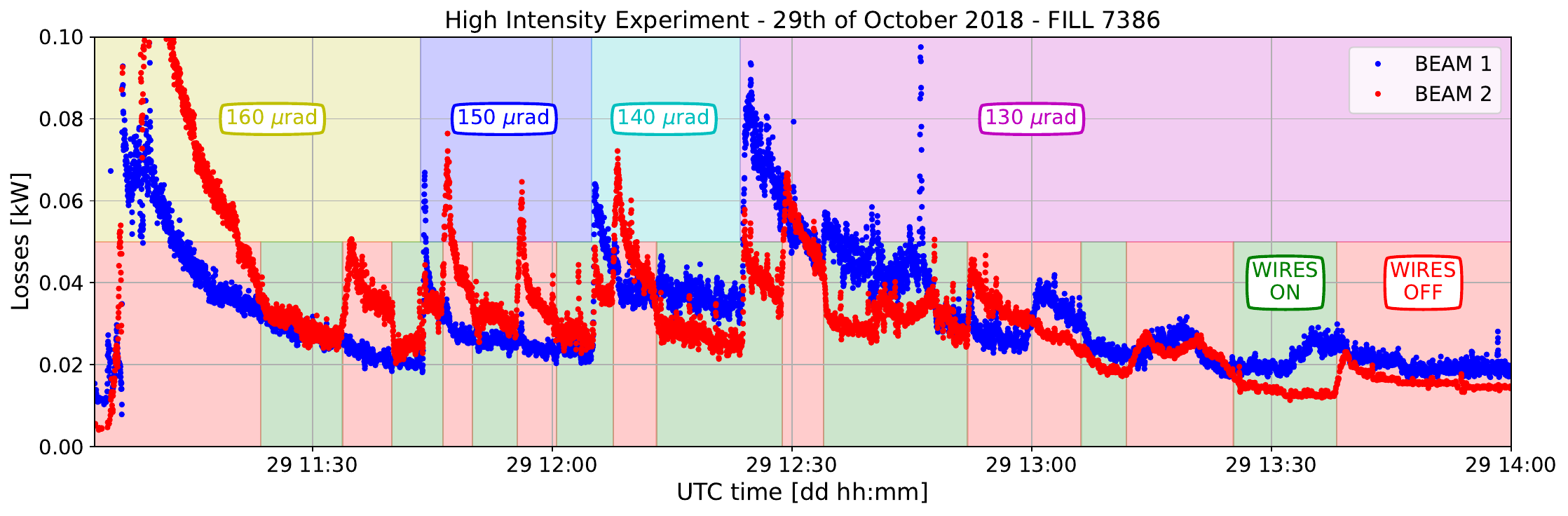} 
\caption{\label{fig:HI_results} High Intensity experiment results. The plot shows the evolution of the BLM losses from Beam 1 (blue dots) and Beam 2 (red dots). One can also see the evolution of the crossing angle on the top background while the bottom background represents the ON (green) and OFF (red) cycles of the wires.}
\end{figure*}

With these settings, the effect of the wires on the bunch-by-bunch effective cross-section was too weak to be clearly observed. Consequently, the adopted observable for the High Intensity experiment was the beam losses recorded from the BLM, which are more sensitive devices. Figure~\ref{fig:HI_results} shows the evolution of the beam losses for both beams (blue and red dots for Beam 1 and Beam 2 respectively), as a function of time. The background coloring correspond to the change of crossing angle and to the powering (on and off in green and red respectively) of the wires.

As in the Low Intensity experiment, the crossing angle was first set to 160~µrad, and the BBLR signature was observed by increasing the transverse beam size of B2, through a controlled emittance blow-up. After some minutes of observation, the wires were turned on. At around 11:35, the wires were turned off and a spike of losses was immediately observed on B2 (as B1 is not equipped with wires). When powering back the wires, the losses decreased to reach the same level as B1, showing that with a crossing angle of 160~µrad the wire compensators were efficient. The crossing angle was then reduced with the wires on. On every step, one can observe a slight spike of losses on B2. Nevertheless, this spike is smaller than the one observed on B1, bringing an evidence of the efficiency of the wire compensators with a configuration compatible with the nominal LHC operation. BBLR interactions mitigation using wires could reduce the beam losses by about 20~\% when reducing the crossing angle step by step.

\begin{figure*}[!htbp]
\includegraphics[width = 2\columnwidth]{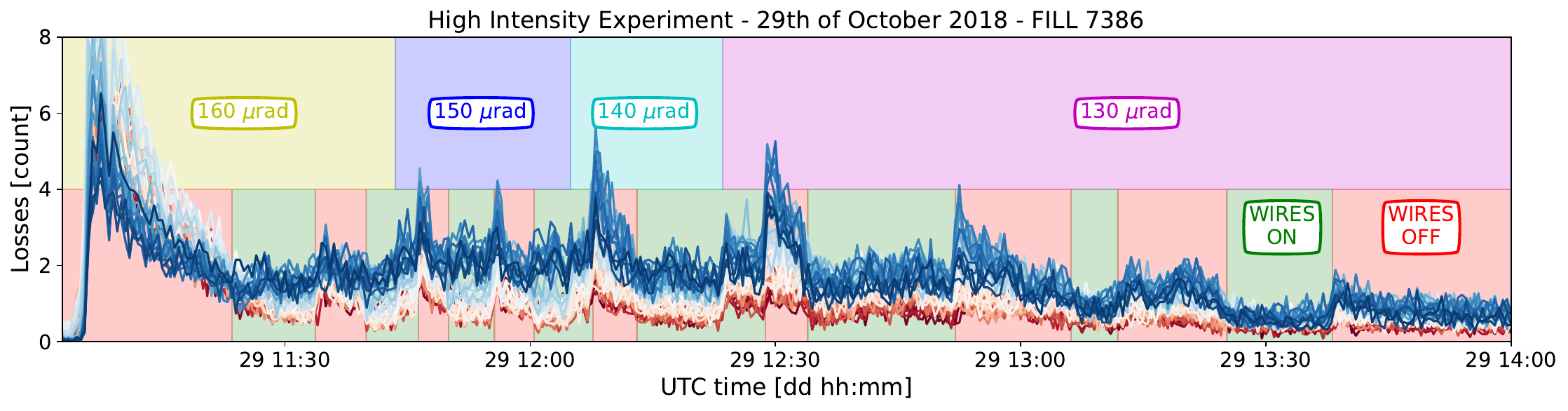} 
\caption{\label{fig:HI_dblm} High Intensity experiment: bunch-by-bunch losses from dBLMs. The background of the plot is colored to show the evolution of the crossing angle and the on/off cycle of the wires. Only the losses from the first train are represented (48 bunches). The color corresponds to the position of the bunch within the train: from red for the first bunch of the train, to blue for the last.}
\end{figure*}

Finally one can also look at the signal provided by the diamond BLMs as these devices provide bunch-by-bunch losses. Figure~\ref{fig:HI_dblm} shows the evolution of the bunch-by-bunch losses in time, together with the powering of the wires and the crossing angle evolution as background colouring. The losses of all the bunches within the train are displayed, and the color depends on the position of the bunch, going from red for the first bunch of the train, to blue for the last. 

The beam losses pattern is quite similar to the one observed on the BLMs: a spike of losses appears when the wires are turned off and can be partially recovered when powering the wires on. Nonetheless, one can extract an additional information from the diamond BLMs: a pattern in the beam losses with respect to the bunch position within the train is observed. Figure~\ref{fig:HI_dblm} shows the dBLM signal acquired during the High Intensity experiment: the losses are shown for different bunch position within the train (from red for the first bunch to blue for the last one). It illustrates that the first bunch of the train always loses less than the last. This cannot be identified as a signature of the BBLR interactions as in that case the central bunch would be the one losing the most. Most likely, this pattern is related to the interplay with electron clouds effects. In fact it is known that the LHC is currently dominated by such effects \cite{Zimmermann:585562, Paraschou:2696125}. The wires still reduce the losses by extending the available tune space, therefore increasing the margins to accommodate extra non-linear effects such as electron clouds, as already shown in the simulations described in Section~\ref{sec:simulations}.

\section{Conclusions and perspectives}

During the 2017-2018 LHC Run, BBLR wire compensators have been built, installed and tested through a diverse experimental campaign. The preparation of this set of experiments was inspired by the work done by S.~Fartoukh \cite{Fartoukh:2052448} based on a resonance driving terms compensation, but also on previous analytical and numerical work showing that wires could compress the tune footprint by compensating the octupolar component of the BBLR interactions \cite{Papaphilippou:574079, PhysRevSTAB.2.104001, Shiltsev1999CompensationOB}. The main goal was to prove that wire compensators could mitigate the BBLR interactions effect in different configurations.

In a first set of experiments, the goal was to provide a proof-of-concept of the BBLR mitigation using DC wires. The wire collimators were brought closer to the beam and, consequently, a low intensity beam - composed of only two bunches - was used. After identifying a clear BBLR signature in terms of beam losses, the wires were turned on and off repeatedly, while reducing the crossing angle. The results of this proof-of-concept showed, for the first time in a high energy collider, in this configuration (colliding beams, in-vacuum wires, inherent BBLR interactions), a clear evidence of a mitigation of the BBLR interactions using wires and motivated a second set of experiments with a setup closer to the nominal LHC operation. 

In this second set of experiments, the goal was to prove the feasibility of the compensation in a configuration compatible with the nominal operation of the LHC. The wire collimators were opened to their nominal settings values and both beams could be composed of several trains of bunches. In order to increase the effective strength of the wire compensators - and in particular, the octupolar component of the field - the two wires housed in the collimators were powered in series. Results showed a possible reduction of beam losses by powering the wires, mainly at lower crossing angle. 

Those positive results yielded a layout change during the Long Shutdown 2, preparatory step for the last LHC Run 3 before the HL-LHC era. It has, in fact, been decided to equip Beam 1 with the two wire collimators installed on Beam 2, located downstream of the IP and that proved to be less effective. The obtained results, together with an important work on simulations, brought a new vision on the available possibilities given the installed hardware. In 2017 those demonstrators were installed in order to prove that a mitigation of the BBLR effect was possible. It is now proposed to power the wires routinely at the end of each fill during the next LHC Run 3 in order to gain experience in operating the machine with such devices, in view of the HL-LHC era.

On the longer term, space is now reserved for the wire compensators in the HL-LHC and they are considered as a complementary solution to the crab cavities since the beam-beam separation, together with the doubled bunch intensity, do not prevent from residual BBLR interactions. The wire compensators could recover - at least, partially - the consequent loss of luminosity \cite{Skoufaris:2777349}. 

Finally this experimental campaign also showed the limitations of the current hardware. The \textit{in-jaw} configuration of the wires should be avoided in the future, as collimator tanks are bulky and can limit the movement of the wires and the additional distance between the wire and the edge of the jaw adds up with the already important beam-wire distance, reducing the efficiency of the compensators.

\section*{Acknowledgments}

The authors of this article would like to thank all the people involved in the wire compensators experimental campaign, in particular: D.~Amorim, G.~Arduini, H.~Bartosik, R.~Bruce, X.~Buffat, L.~Carver, G.~Cattenoz, E.~Effinger, M.~Fitterer, M.~Gasior, M.~Gonzales-Berges, A.~Gorzawski, G.-H.~Hemelsoet, M.~Hostettler, G.~Iodarola, R.~Jones, D.~Kaltchev, S.~Kostoglou, I.~Lamas-Garcia, T.~Levens, A.~Levichev, L.E.~Medina-Madrano, D.~Mirarchi, J.~Olexa, P.S.~Papadopoulou, D.~Pellegrini, L.~Ponce, B.M.~Salvachua Ferrando, H.~Schmickler, F.~Schmidt, R.~Tomás-Garcia, G.~Trad, A.~Valishev, D.~Valuch, C.~Xu, C.~Zamantzas and P.~Zisopoulos. 

\appendix

\section{Alignment of the wire collimators}
\label{app:align}

Each collimator jaw is equipped with two motors allowing a lateral movement of each extremity. These motors are used to control the collimator gap, or to introduce a tilt in order to align the collimator to the beam. Moreover the collimator is equipped with an additional motor, enabling the overall tank to be displaced in the plane orthogonal to the one of collimation. The presence of this so-called 5$^{th}$-axis is motivated by the necessity to offset damaged portions of the jaw in case of beam impact, by an undamaged portion with a simple translation of the  \cite{Redaelli:2010oqa, Valentino:2017mvj}. The wires are parallel to the jaw surface, so their position is controlled by aligning precisely the collimator to the beam. During the nominal operation of the machine, the beam orbit is centered in the middle of the two jaws. However, several constraints might occur depending on the chosen crossing scheme. The wire alignment (both transverse and 5$^{th}$-axis) should therefore always be checked.

The alignment of the LHC collimators has been improved by the installation of Beam Position Monitors (BPM) \cite{Garcia-TabaresValdivieso:2701414} at their entrance and their exit \cite{WOLLMANN201462}. Differently from the jaw alignment, the 5$^{th}$-axis alignment requires a scan of its position. The BPMs embedded in the collimators give the absolute position of the beam in between the jaws, but there is not an equivalent BPM for the 5$^{th}$-axis. To proceed with this alignment one can monitor the signal of those BPMs while moving the collimator tank. If the 5$^{th}$-axis is misaligned, the signal recorded by the BPMs becomes weaker. The intensity of this signal is linear with the beam intensity (a normalization of the signal by the beam intensity might therefore be needed) while the misalignment is a second order effect with respect to the jaw position. For small variations around the aligned position, the BPM reading as a function of the 5$^{th}$-axis position is fitted well by a parabola whose maximum corresponds to the position of the motors that aligns the collimator - and the wire - to the beam.

\begin{figure}[!htbp]
\includegraphics[width = \columnwidth]{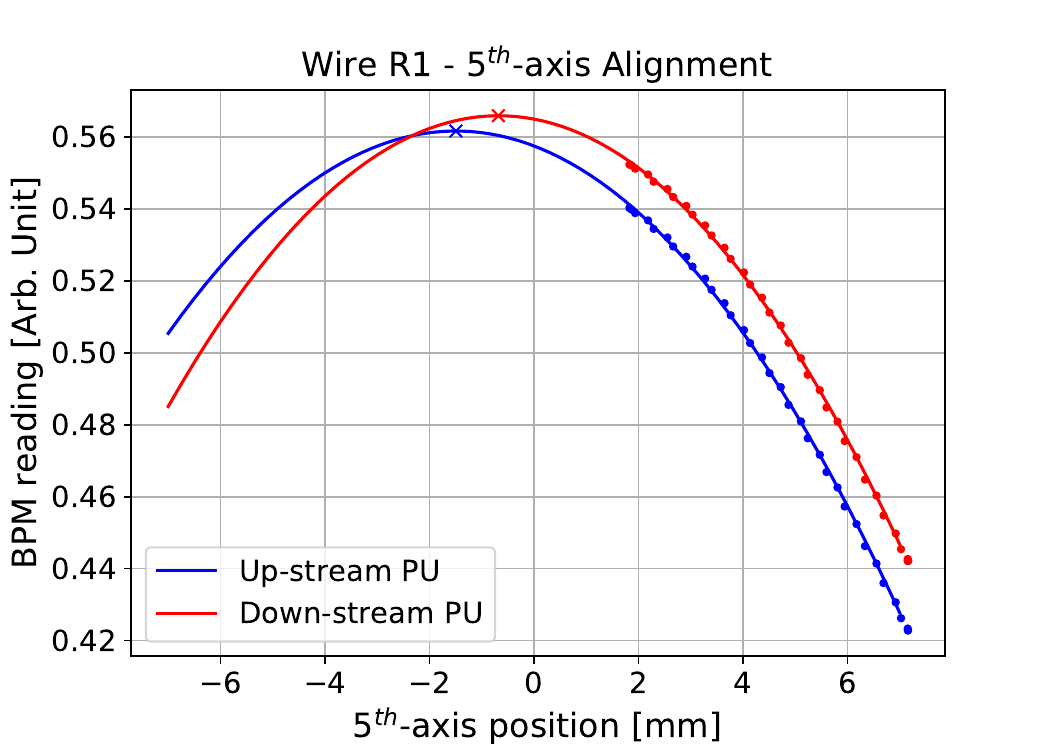} 
\caption{\label{fig:align} 5$^{th}$-axis collimator (Wire R1) alignment. The dot lines represent the measurement while the solid lines are the second order polynomial fits. The maximum of each parabola is indicated by a cross.}
\end{figure}

Figure~\ref{fig:align} gives an example of the alignment of the wire collimator on the right side of IP1. As one can see from the BPM signal, the maximum of the parabola could not be reached: the collimator motion was blocked by the vacuum chamber of the opposite beam. Fitting the available data (red and blue dots) with a second order polynomial (red and blue lines) allowed us to determined the misalignment and to intervene in the tunnel in order to recover it. 

This example illustrated the limitations of the in-jaw configuration of the wires and the need for a local beam diagnosis for a fast and precise alignment. 

\section{BBLR compensation and $\beta$-beating}
\label{app:beta_beat}

$\beta$-beating in synchrotrons and colliders is a consequence of a perturbation, resulting in the modification of the $\beta$-functions all along the machine. A local quadrupolar perturbation can induce a modification of the $\beta$-function that would then propagate under the form of a wave along the machine, following the betatronic motion. Effects such as beam-beam interactions (or their compensation using DC wires) can be the origin of such perturbations and must be taken into consideration. The source of a $\beta$-beating wave is quadrupolar but does not necessarily originate from quadrupole errors as other higher order magnet errors can induce feed-down effects.

If the source of the perturbation is known, one can compute analytically the $\beta$-beating wave according to Eq.~\ref{eq:beta_beat_analy} \cite{handbook_error_sources}:
\begin{align}
\begin{split}
\label{eq:beta_beat_analy}
\frac{\Delta \beta_{x,y} (s)}{\beta_{x,y}(s)} = \frac{1}{2 \sin(2 \pi Q_{x,y})} \sum_i (\Delta K l)_i \beta_{x,y}(s_i) \\ \times \cos(2|\phi_{x,y}(s) - \phi_{x,y}(s_i)| - 2\pi Q_{x,y}),
\end{split}
\end{align}
where $Q_{x,y}$ stands for the horizontal or vertical betatron tunes, $(\Delta K l)_i$ represents the $i^{th}$ perturbation source located at a position $s_i$ and $\phi_{x,y}$ the horizontal or vertical phase advance. 

Another way to access information about the $\beta$-beating in a case where it would be difficult to compute it analytically is the use of the MAD-X code \cite{madx}, where one can use the TWISS function before and after the application of the perturbation to observe the effect on the $\beta$-functions. 

Finally, it is also possible - even though more difficult - to measure the $\beta$-functions experimentally. An example of $\beta$-beating measurement in the LHC can be found in \cite{Boccardi:1272165}. 

\begin{figure*}[!htbp]
\includegraphics[width = 2\columnwidth]{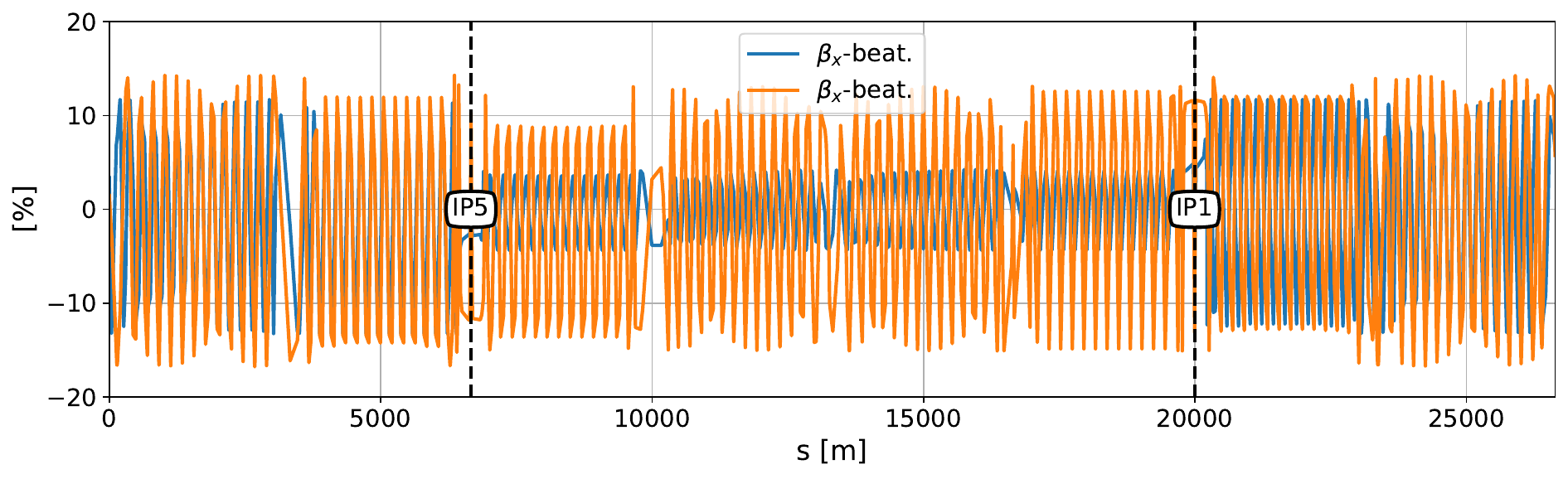} 
\caption{\label{fig:beta_beat_match} $\beta$-beating induced by the wire compensators, with the tunes matched using the tune correctors instead of the feed-forward system.}
\end{figure*}

In the case of BBLR interactions and their mitigation using the wire compensators (and their associated feed-forward system described in Section~\ref{sec:FF}), a study has been carried out using the MAD-X code. Firstly, one can study the $\beta$-beating induced by the wire compensators alone, in the case of a tune matching using the nominal tune correctors of the machine, as shown in Fig.~\ref{fig:beta_beat_match}.

\begin{figure*}[!htbp]
\includegraphics[width = 2\columnwidth]{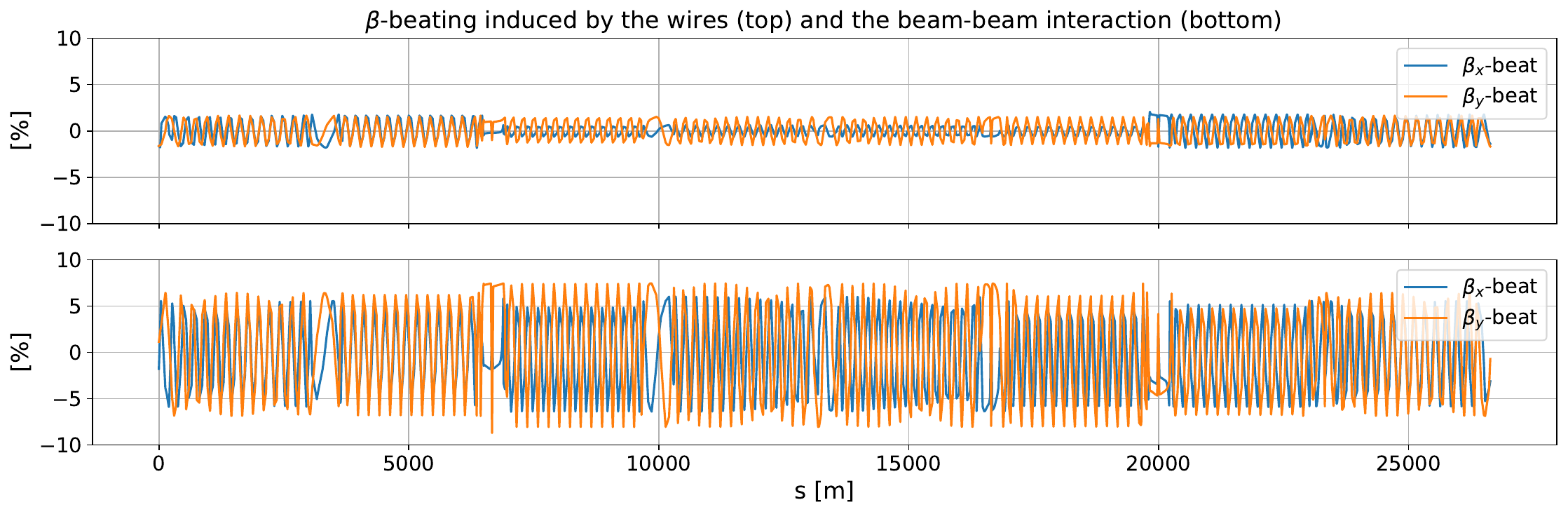} 
\caption{\label{fig:beta_beat} Example of a $\beta$-beating induced by the beam-beam interactions and the wire compensators (together with their feed-forward system). On the top plot, the beatings in the horizontal and vertical plane are represented as a function of the longitudinal position, in presence of the wires but without any beam-beam interactions (no HO neither BBLR). The bottom plot represents the same beating in presence of the wires and the beam-beam interaction. The chosen optics in this example is close to the one used in the High Intensity experiment (half crossing angle of 150~µrad, $\beta^{*}$ of 30 cm). The wires are powered with their maximum current and the 2-jaw powering configuration is used (see Sections~\ref{sec:exp_protocol} and \ref{sec:exp_results}).}
\end{figure*}

As one can see in Fig.~\ref{fig:beta_beat_match}, using the tunes correctors to compensate the tune shifts induced by the wire compensators yields an important $\beta$-beating of more than 10~\% along the machine. However, in Fig.~\ref{fig:beta_beat}, an example of $\beta$-beating induced by the wire compensators in the case of the feed-forwards, with a comparison to the $\beta$-beating induced by the beam-beam interactions, can be observed. 

It can be observed on the top plot that the contribution of the wires to the $\beta$-beating is at the level of around 1~\% while the beam-beam interactions (BBLR and HO together) bring the $\beta$-beating level along the machine around 10~\%. For that reason, the $\beta$-beating induced by the wires and their feed-forward system is considered to be negligible.

\section{Analytical computation of the optimal wire currents}
\label{app:rdt_comp}

In this Appendix we report the method used in order to determine the optimal wire currents. The approach is based on \cite{Fartoukh:2052448} from which we recall the main concepts. 

The Resonance Driving Terms (RDT) of order $n = p + q$ driven by the BBLR interactions are denoted $c_{p,q}^{LR}$ and are defined in Eq.~\ref{eq:cpq_lr}:
\begin{align}
\label{eq:cpq_lr}
    c_{p,q}^{LR} = \sum_{k \in LR} \frac{\beta_x^{p/2}(s_k)\beta_y^{q/2}(s_k)}{d_{bb}^{p+q}},
\end{align}
where $\beta_{x,y}(s_k)$ are the horizontal or vertical $\beta$-functions at the $k^{th}$ BBLR interaction, located at $s=s_k$ and $d_{bb}$ is the beam-beam separation. Similarly one can define the RDTs led by a wire compensator as in Eq.~\ref{eq:cpq_w}:
\begin{align}
\label{eq:cpq_w}
    c_{p,q}^{w} = N_w \frac{\beta_{x,w}^{p/2}\beta_{y,w}^{q/2}(s_k)}{d_{w}^{p+q}},
\end{align}
where $N_w$ is the integrated current of the wire expressed in terms of equivalent number of BBLR interactions, $\beta_{x,y,w}$ the $\beta$-functions at the wire location and $d_{w}$ the beam-wire distance. 

Considering each IP independently, and adding the subscripts $L$ and $R$ for the left and right wires respectively, one has to solve the system of Eqs.~\ref{eq:sys_rdt} in order to determine the current required in each wire to compensate the ($p_1,q_1$)-($p_2,q_2$) RDTs:
\begin{subequations}
  \begin{empheq}[left=\empheqlbrace]{align}
    c_{p_1,q_1}^{w,L} + c_{p_1,q_1}^{w,R} &= - c_{p_1,q_1}^{LR} \\
    c_{p_2,q_2}^{w,L} + c_{p_2,q_2}^{w,R} &= - c_{p_2,q_2}^{LR}.
  \end{empheq}
  \label{eq:sys_rdt}
\end{subequations}

Solving this system for the left and right wire currents $N_{w,L}$ and $N_{w,R}$, one can finally obtain:
\begin{widetext}
\begin{align}
    N_{w,L} &= \frac{d_L^{p_1+q_1+p_2+q_2}(c_{p_1,q_1}^{LR}d_R^{p_1+q_1}\beta_{x,R}^{p_2/2}\beta_{y,R}^{q_2/2}-c_{p_2,q_2}^{LR}d_R^{p_2+q_2}\beta_{x,R}^{p_1/2}\beta_{y,R}^{q_1/2})}{d_L^{p_2+q_2}d_R^{p_1+q_1}\beta_{x,L}^{p_1/2}\beta_{x,R}^{p_2/2}\beta_{y,L}^{q_1/2}\beta_{y,R}^{q_2/2}-d_L^{p_1+q_1}d_R^{p_2+q_2}\beta_{x,L}^{p_2/2}\beta_{x,R}^{p_1/2}\beta_{y,L}^{q_2/2}\beta_{y,R}^{q_1/2}}\\
    N_{w,R} &= \frac{d_R^{p_1+q_1+p_2+q_2}(c_{p_1,q_1}^{LR}d_L^{p_1+q_1}\beta_{x,L}^{p_2/2}\beta_{y,L}^{q_2/2}-c_{p_2,q_2}^{LR}d_L^{p_2+q_2}\beta_{x,L}^{p_1/2}\beta_{y,L}^{q_1/2})}{d_L^{p_1+q_1}d_R^{p_2+q_2}\beta_{x,L}^{p_2/2}\beta_{x,R}^{p_1/2}\beta_{y,L}^{q_2/2}\beta_{y,R}^{q_1/2}-d_L^{p_2+q_2}d_R^{p_1+q_1}\beta_{x,L}^{p_1/2}\beta_{x,R}^{p_2/2}\beta_{y,L}^{q_1/2}\beta_{y,R}^{q_2/2}}.
\end{align}
\end{widetext}

From these expressions one can compute the needed currents to compensate for instance the (4,0)-(0,4) RDT, as described in Section~\ref{sec:wire_settings} and Fig.~\ref{fig:opt_currents}.

\bibliography{biblio}

\end{document}